\documentclass[bibyear]{aa} 
\usepackage{xcolor}
\usepackage{longtable}
\usepackage{multirow}
\usepackage{adjustbox} 
\usepackage{gensymb}
\usepackage{tabularx} 
\usepackage{amsmath}
\usepackage{natbib}
\usepackage{pdflscape}
\usepackage{lscape}
\usepackage{ulem}
\usepackage{graphicx}
\usepackage{txfonts}
\usepackage{hyperref}
\titlerunning{High-frequency humps in \Swift} 
\authorrunning{Ze-Xi Li et al.}
\begin{document} 

\newcommand{\Swift}{Swift J1727.8--1613}
\newcommand{\HXMT}{$Insight$-HXMT}
   \title{Characteristics of the High-frequency Humps in the Black hole X-ray Binary \Swift}

   \author{Ze-Xi Li\inst{1}\fnmsep\inst{2}, Liang Zhang\inst{1}, Lian Tao\inst{1}, Zi-Han Yang\inst{1}\fnmsep\inst{2}, Qing-Chang Zhao\inst{1}\fnmsep\inst{2}, Shu-Jie Zhao\inst{1}\fnmsep\inst{2}, Rui-Can Ma\inst{3}, Zi-Xu Yang\inst{4}, Pan-Ping Li\inst{1}\fnmsep\inst{2}, Xiang Ma\inst{1}, Yue Huang\inst{1}, Shu-Mei Jia\inst{1}, Shuang-Nan Zhang\inst{1}\fnmsep\inst{2}\fnmsep\inst{5}, Hua Feng\inst{1}, Jin-Lu Qu\inst{1}
          \and 
          Shu Zhang\inst{1}
          }

   \institute{Key Laboratory for Particle Astrophysics, Institute of High Energy Physics, Chinese Academy of Sciences, 19B Yuquan Road, Beijing 100049, China \\
              \email{zhangliang@ihep.ac.cn, taolian@ihep.ac.cn}
         \and
            University of Chinese Academy of Sciences, Chinese Academy of Sciences, Beijing 100049, China
        \and
            Department of Physics and Astronomy, University of Southampton, Highfield, Southampton, SO17 1BJ, UK
        \and
            School of Physics and Optoelectronic Engineering, Shandong University of Technology, Zibo 255000, China
        \and
            Key Laboratory of Space Astronomy and Technology, National Astronomical Observatories, Chinese Academy of Sciences, Beijing 100101, China
             }


 
    \abstract{
    We present a detailed timing analysis of the two high-frequency humps observed in the power density spectrum of \Swift\ up to 100\,keV, using data from the Hard X-ray Modulation Telescope (\HXMT). 
    Our analysis reveals that the characteristic frequencies of the humps increase with energy up to $\sim$30\,keV, followed by a plateau at higher energies. The fractional rms amplitudes of the humps increase with energy, reaching approximately 15\% in the 50-100\,keV band. The lag spectrum of the hump is characterized primarily by a soft lag that varies with energy.
    Our results suggest that the high-frequency humps originate from a corona close to the black hole.
    Additionally, by applying the relativistic precession model, we constrain the mass of \Swift\ to $2.84 < M / M_{\odot} < 120.01$ and the spin to $0.14 < a < 0.43$ from the full-energy band dataset, using triplets composed of a type-C quasi-periodic oscillation and two high-frequency humps.  When considering only the high-energy bands with stable characteristic frequencies, we derive additional constraints of $2.84 < M/M_{\odot} < 13.98$ and $0.14 < a < 0.40$.
    %
    }


   \maketitle

%
\section{Introduction}
    Black hole low-mass X-ray binaries (BH-LMXBs) in outbursts usually exhibit fast X-ray aperiodic variability on a wide range of timescales \citep[see reviews by][]{2006ARA&A..44...49R, 2007A&ARv..15....1D}. Fourier analysis serves as an effective and intuitive approach for examining these complex timing variations. In a typical power density spectrum (PDS) of BH-LMXBs, several narrow peaks, known as quasi-periodic oscillations (QPOs), are commonly observed alongside a broad-band noise (BBN) continuum \citep[see reviews by][]{Belloni2016, 2019NewAR..8501524I}.
    Based on their frequencies, QPOs can be categorized into two main groups: low-frequency QPOs (LFQPOs, e.g. \citealt{1979Natur.278..434S, 1985Natur.316..225V, 1995ApJ...441..779V, 1999ApJ...526L..33W,Homan2001,2002ApJ...564..962R, 2005ApJ...629..403C, Motta2011}), which have frequencies ranging from 0.1 to 30\,Hz, and high-frequency QPOs (HFQPOs, e.g. \citealt{1997ApJ...482..993M, 1999ApJ...522..397R, 1999ApJ...520..262P, 2001ApJ...552L..49S, 2012MNRAS.426.1701B, 2013MNRAS.435.2132M}), typically occurring in the range of 40 to 450\,Hz.
    
    LFQPOs have been observed in almost all BH-LMXBs. Based on the shape of the PDS and the spectral state during which the QPOs are detected, LFQPOs can be further categorized into types A, B, and C \citep{2002ApJ...564..962R, 2005ApJ...629..403C}. 
    The origins of LFQPOs are primarily attributed to two classes of models: accretion disk instability models \citep{1990PASJ...42...99K,  1996ApJ...457..805M, 1999A&A...349.1003T} and geometrical effect models. The latter includes, for example, Lense-Thirring precession of the hot inner flow \citep{2009MNRAS.397L.101I} or the precession of the jet base \citep{2016MNRAS.460.2796S, 2021NatAs...5...94M}. 
    Evidence has been provided suggesting that type-C QPOs may have a geometric origin, as indicated by the inclination-dependent fractional rms of QPOs and the modulation of the iron line with QPO phase \citep{2015MNRAS.447.2059M, 2016MNRAS.461.1967I}. However, models related to geometrical effects still encounter considerable challenges, both theoretically and observationally \citep{2021ApJ...906..106M, 2024ApJ...961L..42Z}.
    
    HFQPOs have been observed in only a limited number of BH-LMXBs \citep[e.g.,][]{2001ApJ...552L..49S, 2012MNRAS.426.1701B, 2013MNRAS.435.2132M}. 
    The frequency of the HFQPOs usually occurs at specific values and does not change significantly with luminosity \citep{2002ApJ...580.1030R, 2006ApJ...637.1002R}. 
    In some sources, double peaks have been observed, with the frequency ratio consistently approximating 3:2 \citep{ 2001ApJ...552L..49S, 2002ApJ...580.1030R, 2005ApJ...623..383H}. 
    Furthermore, HFQPOs are generally very weak, with fractional rms amplitudes typically below 5\% \citep{2012MNRAS.426.1701B}.
    Numerous theoretical models have been proposed to explain the underlying mechanisms of HFQPOs, although their physical origin remains highly debated. 
    \cite{2001A&A...374L..19A} proposed a resonant mechanism, specifically accounting for the observed 3:2 frequency ratio. However, the robustness of this frequency ratio remains uncertain due to the limited number of detections.
    Another widely used model to explain HFQPOs is the relativistic precession model (RPM, \citealt{1998ApJ...492L..59S, 1999PhRvL..82...17S, 1999ApJ...524L..63S}). 
    The RPM associates the nodal precession frequency ($\nu_{\mathrm{nod}}$), the periastron precession frequency ($\nu_{\mathrm{per}}$), and the orbital frequency ($\nu_{\phi}$) at the same radius with the type-C QPO, the lower HFQPO, and the upper HFQPO frequency, respectively.
    When HFQPO pairs and a type-C QPO are observed simultaneously, the mass and spin of the black hole can be determined analytically using the RPM.
    This approach has been successfully applied to estimate the mass and spin parameters of several BH-LMXBs \citep{2014MNRAS.437.2554M, 2014MNRAS.439L..65M, 2019MNRAS.486.4485D,2022MNRAS.517.1469M}. 

    Among the components of BBN, the high-frequency humps, characterized by frequencies exceeding approximately 30\,Hz, 
    are generally the highest frequency features detected in the PDS of BH-LMXBs. These humps play a critical role in investigating the short-timescale dynamics of matter in the innermost regions around black holes. Such humps have been observed in numerous sources \citep{2000MNRAS.318..361N,2001ApJ...558..276T,2002ApJ...572..392B,2003ApJ...586..419K,2014MNRAS.437.2554M,2014MNRAS.439L..65M,2021MNRAS.508.3104B,2022MNRAS.514.2839A,2022MNRAS.517.1469M}.
    The characteristic frequencies of the humps exhibit a strong correlation with the frequency of the LFQPO \citep{1999ApJ...520..262P, 2025arXiv250303078F}. The fractional rms amplitudes of the humps typically increase with photon energy \citep{2022MNRAS.514.2891Z,2024MNRAS.527.5638Z}. 
    In the case of GRS 1915+105, the rms amplitudes of the humps show a positive correlation with the corona temperature and an anti-correlation with the radio flux \citep{2022NatAs...6..577M,2022MNRAS.514.2891Z}. 
    This suggests that the humps can serve as an indicator of accretion energy distribution. When the humps are strong, the majority of the accretion energy is channeled into the corona rather than being transferred to the jet.
    \cite{2022MNRAS.514.2891Z} further proposed that the humps and the HFQPO in GRS 1915+105 originate from the same variability component, with the coherence of this component being determined by the properties of the corona.
    
    

    The bright new X-ray transient \Swift\ was discovered by Swift/BAT on 2023 August 24 \citep{2023GCN.34544....1N, 2023GCN.34540....1K}.
    Dynamical measurements have confirmed the compact object to be a black hole with a mass of $M > 3.12 \pm 0.10\,M_{\odot}$ \citep{2025A&A...693A.129M}.
    An extremely high spin value of 0.98 was obtained from the reflection modeling \citep{2024arXiv240603834L}.
    The distance to \Swift\ was initially estimated to be $d=2.7\pm 0.3$ kpc using several empirical methods \citep{2024A&A...682L...1M}. However, \citealp{2025arXiv250206448B} suggested an increased distance of $5.5^{+1.4}_{-1.1}$ kpc.
    Additionally, \cite{2024ApJ...971L...9W} identified a bright core and a large, two-sided, asymmetrical, resolved jet using the VLBA and LBA observations. They constrained the jet speed to $\beta \geq 0.27$ and the jet inclination to $i\leq 74^\circ$.


    \citet{2024MNRAS.529.4624Y} conducted a comprehensive analysis of the evolution of the X-ray variability in \Swift\ using \HXMT\ observations. They detected prominent type-C QPO with frequencies ranging from 0.1\,Hz to 8\,Hz. Additionally, significant high-frequency humps were observed in the PDS, whose frequencies are highly correlated with the QPO frequencies.
    In this study, we present a detailed investigation of the evolution of the high-frequency humps and their energy-dependent properties up to 100\,keV using \HXMT\ observations. Furthermore, we determine the black hole mass and spin in \Swift\ by applying the RPM.
    In Section \ref{section 2}, we introduce our data selection and reduction methodology. We present our data analysis and results in Section \ref{section 3}. In Section \ref{section 4}, we discuss our main findings.

\section{Observation and Data Reduction}
\label{section 2}

    \begin{figure*}
    \centering
    \includegraphics[width=1\linewidth]{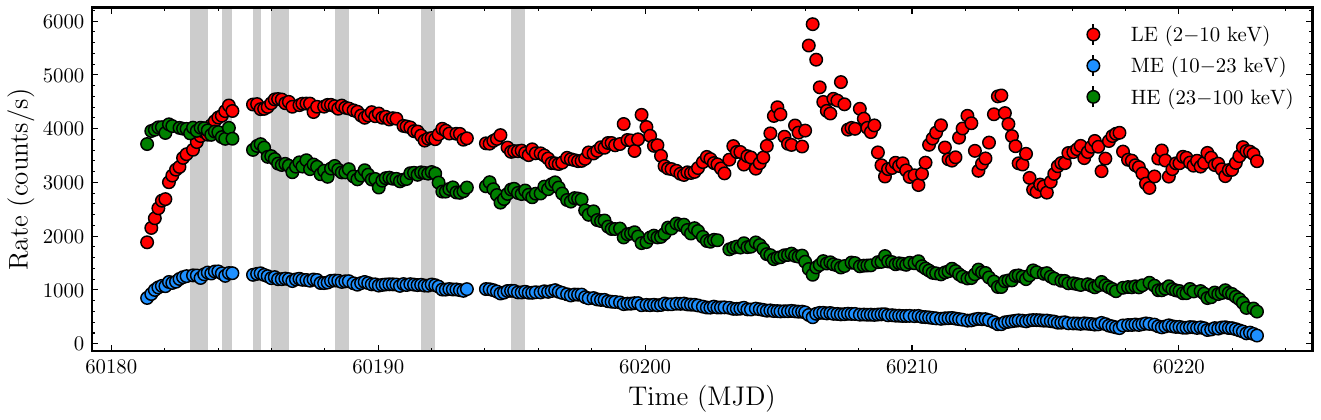}
    \caption{\HXMT\ LE 2-10\,keV, ME 10-23\,keV, and HE 23-100\,keV light curves of \Swift\ during its 2023 outburst. Each data point corresponds to an exposure ID. The shaded regions mark the data groups selected for our timing analysis in this work.}
            \label{Fig1}%
    \end{figure*}
    \begin{figure}
    \centering
    \includegraphics[width=\hsize]{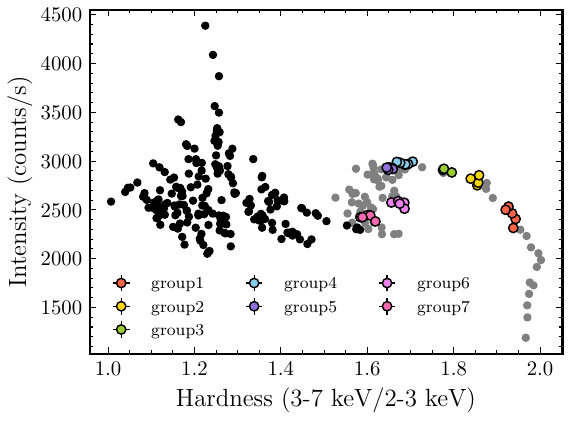}
        \caption{\HXMT\ hardness-intensity diagram (HID) of \Swift\ during its 2023 outburst. Each data point corresponds to an exposure ID. The different colored points represent the data groups selected for our timing analysis in this work. }
            \label{Fig2}
    \end{figure}

    \begin{table*}
        \centering
        \caption[]{Log of the 7 data groups selected for our timing analysis in this work. Only the last five digits of the exposure IDs are shown, with the common prefix P06143380 omitted for brevity.}
            \label{Table1}
            \resizebox{\textwidth}{!}{
            \begin{tabular}{ccccccc}
            
                \hline
                \noalign{\smallskip}
                Exposure ID &  Time & 2-10\,keV Rate  & 10-23\,keV Rate & 23-100\,keV Rate & Hardness & Type-C QPO frequency\\
                 & (MJD) & (counts $\mathrm{s^{-1}})$ & (counts $\mathrm{s^{-1}})$ & (counts $\mathrm{s^{-1}})$ &  & (Hz)\\
                \noalign{\smallskip}
                \hline
                \noalign{\smallskip}
                00113 - 00205  &  60182.95 - 60183.61  &  
                $3810^{+130}_{-200}$  &  
                $1280 \pm 50$  &  
                $3980^{+40}_{-70}$  &  
                $1.933^ {+0.011}_{-0.012}$  &
                $0.50^{+0.03}_{-0.04}$\\
                \noalign{\smallskip}
                \hline
                \noalign{\smallskip}
                00209 - 00212  &  60184.14 - 60184.54 &  
                $4330^{+100}_{-90}$  &  
                $1300^{+20}_{-40}$  &  
                $3800^{+140}_{-60}$  &  
                $1.853^ {+0.006}_{-0.013}$  &
                $0.700^{+0.011}_{-0.018}$\\
                \noalign{\smallskip}
                \hline
                \noalign{\smallskip}
                00301 - 00303  &  60185.31 - 60185.59  &  
                $4420^{+30}_{-60}$  &  
                $1297^{+13}_{-15}$  &  
                $3660^{+50}_{-60}$  &  
                $1.783^ {+0.013}_{-0.008}$  &
                $0.86^{+0.03}_{-0.05}$\\
                \noalign{\smallskip}
                \hline
                \noalign{\smallskip}
                00306 - 00311 &  60185.99 - 60186.65  &  
                $4510 \pm 40$  &  
                $1212^{+27}_{-12}$  &  
                $3380^{+110}_{-70}$  &  
                $1.69\pm0.02$  &
                $1.12^{+0.03}_{-0.06}$\\
                \noalign{\smallskip}
                \hline
                \noalign{\smallskip}
                00410 - 00414 &  60188.37 - 60188.90  &  
                $4400 \pm 30$  &  
                $1164^{+12}_{-15}$  &  
                $3230^{+70}_{-50}$  &  
                $1.652\pm0.007$  &
                $1.22^{+0.03}_{-0.04}$\\
                \noalign{\smallskip}
                \hline
                \noalign{\smallskip}
                00605 - 00609 &  60191.61 - 60192.14 &  
                $3820^{+50}_{-40}$  &  
                $1090^{+11}_{-10}$  &  
                $3177^{+16}_{-15}$  &  
                $1.675^{+0.011}_{-0.019}$  &
                $1.11^{+0.05}_{-0.03}$\\
                \noalign{\smallskip}
                \hline
                \noalign{\smallskip}
                00808 - 00904 &  60194.98 - 60195.51 &  
                $3573^{+18}_{-27}$  &  
                $969^{+16}_{-15}$  &  
                $2840^{+40}_{-50}$  &  
                $1.60\pm 0.02$  &
                $1.34^{+0.07}_{-0.04}$\\
                \noalign{\smallskip}
                \hline
            \end{tabular}}
    \end{table*}

    \HXMT\ is renowned for its broad energy range, spanning from 1\,keV to 250\,keV, achieved through the utilization of three distinct telescopes: the High Energy X-ray telescope (HE, 20-250\,keV, \citealt{2020SCPMA..6349503L}) , the Medium Energy X-ray telescope (ME, 5-30\,keV, \citealt{2020SCPMA..6349504C}) and the Low Energy X-ray telescope (LE, 1-15\,keV, \citealt{2020SCPMA..6349505C}). 
    
    \HXMT\ observed the outburst of \Swift\ from August 25 to October 4, 2023, with a high cadence over a total duration exceeding one month.
    We processed all the data using the \HXMT\ Data Analysis Software v2.06. The Good Time Interval (GTI) filtering criteria were based on the default standards recommended by the \HXMT\ team: (1) Earth elevation angle larger than $10\degree$; (2) pointing offset angle less than $0.04\degree$; (3) the value of the geomagnetic cutoff rigidity larger than 8 GV; (4) at least 300 s before and after the South Atlantic Anomaly passage. 
    
    In Fig.~\ref{Fig1}, we show the LE 2-10\,keV, ME 10-23\,keV, and HE 23-100\,keV light curves for the outburst of \Swift. The LE light curve exhibits a rapid rise followed by an exponential decay, alongside several flares occurring between MJD 60197 and MJD 60220. Notably, these flares are not observed in the ME and HE light curves.
    In Fig.~\ref{Fig2}, we plot the hardness-intensity diagram (HID). The hardness is defined as the ratio of the count rate between 2-3\,keV and 3-7\,keV energy bands, while the intensity corresponds to the count rate in the 2-7\,keV band. 

   In this study, we focus on the analysis of high-frequency humps, which were only detected in the PDS prior to the flaring state \citep[see][]{2024MNRAS.529.4624Y}.
   For our analysis, we selected 7 data groups by merging several adjacent ExpIDs that exhibit consistent source flux, spectral hardness ratios, and PDS shapes to ensure robust statistical results.
   To achieve this, we applied the following criteria: 
   (1) the variation in the count rate for each detector must remain within 0.05 times the average value\footnote{It is noteworthy that Group 1 corresponds to the initial rapid increase phase of the outburst, during which the LE count rate exhibits significant variations.};
   (2) the variation in the hardness ratio must be less than 0.02;
   (3) the variation in the QPO frequency must remain within 0.1\,Hz.
   The total effective exposure time for each group exceeds 6\,ks.
   A detailed log of the observations analyzed in this study is listed in Tab.\ref{Table1}. The data groups are also marked in Figs.\ref{Fig1} and \ref{Fig2}.

    \begin{figure*}
    \centering
    \includegraphics[width=\hsize]{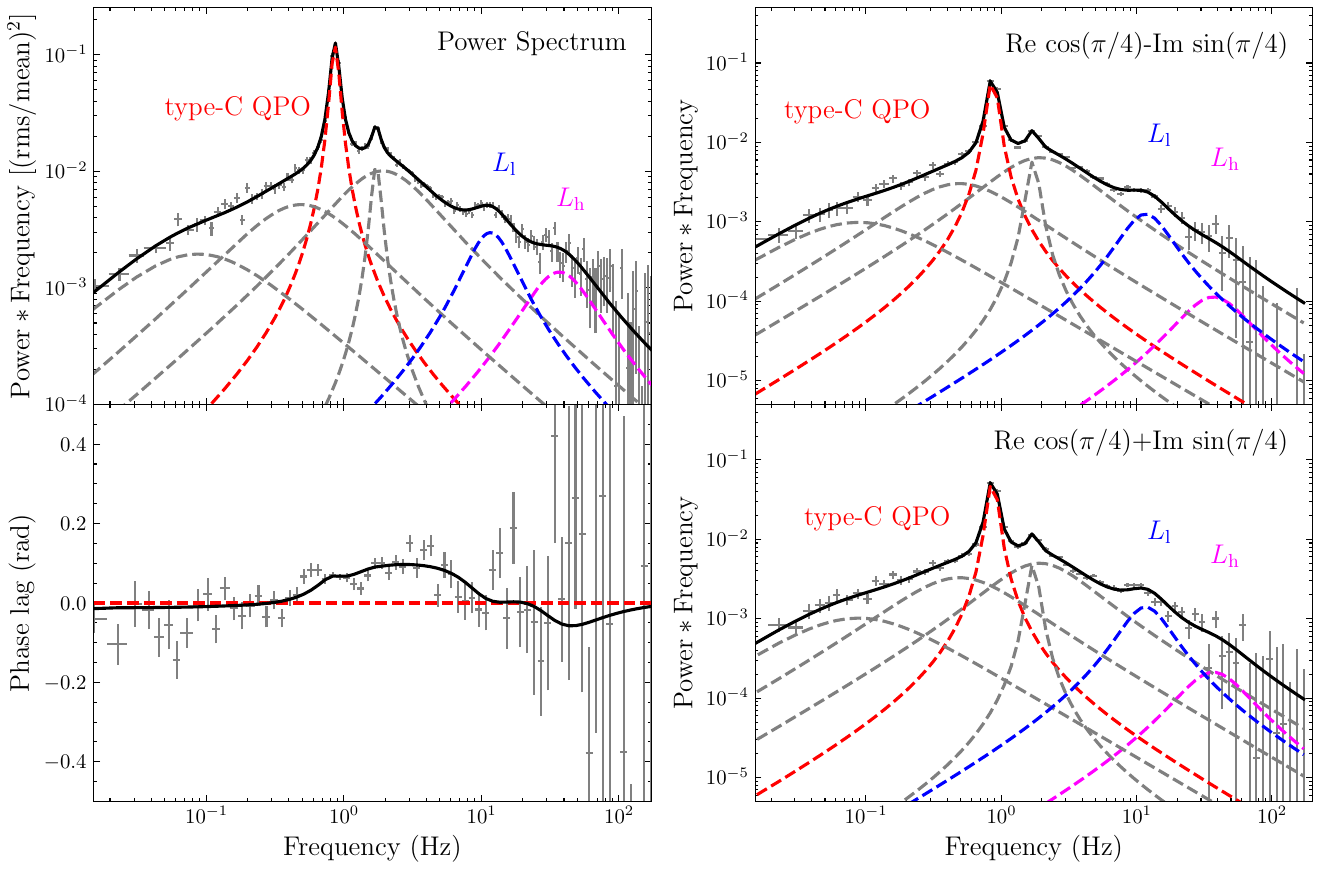}
        \caption{Upper-left panel: PDS in the 4-10\,keV energy band. Lower-left panel: phase lag versus Fourier frequency (phase-lag spectrum) together with the derived model obtained from the fits to the power and cross spectra. Upper-right and lower-right panels:  Real and Imaginary parts of the cross spectrum calculated for the 4-10\,keV band with respect to the 2-4\,keV band. For the fitting and the plot, we rotated the cross-vector by $45\degree$. We modeled the PDS using 7 Lorentzian functions. Additionally, we fitted the Real and Imaginary parts of the cross spectrum by fixing the frequency and FWHM of each Lorentzian to the values derived from the best-fitting model of the PDS. Type-C QPO corresponds to the Lorentzian function used to fit the type-C QPO, while $L_{\rm l}$ and $L_{\rm h}$ represent the Lorentzian functions fitting the two high-frequency humps. } 
            \label{Fig3}
    \end{figure*}
\section{Data Analysis and Results}
\label{section 3}

\subsection{Power density spectrum and cross spectrum}
\label{section 3.1}
    
    For each group, we produced averaged PDS in different energy bands using a time interval of 128 s and a time resolution of 1 ms. The resulting PDS were normalized in units of (rms/mean)$^{2}$~Hz$^{-1}$ \citep{Belloni1990}, and the Poisson noise level estimated from the power between 200 and 500\,Hz was subtracted. To account for background, we applied a correction by multiplying the power by $\left(\frac{S+N}{S}\right)^2$, where $N$ and $S$ represent the count rates of the background and the source, respectively. We modeled the PDS using a combination of multiple Lorentzian functions.

    In the upper-left panel of Fig.~\ref{Fig3}, we show a representative PDS of Group 3, calculated in the 4-10\,keV band. 
    The PDS can be well described by a model comprising seven Lorentzian functions.
    A sharp type-C QPO is prominently observed, accompanied by its second-harmonic peak. 
    The low-frequency noise is fitted using three Lorentzian functions. Additionally, two high-frequency humps are significantly detected, denoted as $L_{\rm l}$ and $L_\mathrm{h}$. We note that the component $L_\mathrm{h}$ was not detected in the PDS analyzed by \citet{2024MNRAS.529.4624Y}, as their study was confined to the frequency range below 50\,Hz.
    

    To determine the lags of the variability components identified in the PDS, we also computed averaged cross spectra between different energy bands using the same time interval and time resolution as those adopted for PDS. 
    %
    %
    Traditionally, the lag of a BBN component or a QPO is measured as the ratio of the average of the real and imaginary parts of the cross spectrum within the selected frequency range (typically $\nu_{0} \pm \rm FWHM/2$; see, e.g., \citealt{1987ApJ...319L..13V}). This method assumes that the component of interest dominates the power and cross spectra over the frequency range of interest.
    However, in cases where other components contribute significantly to the power and cross spectra in the selected frequency range, this method becomes ineffective.
    To measure the lags of weak variability components, \citet{2024MNRAS.527.9405M} introduced a novel method based on simultaneously fitting the PDS and the Real and Imaginary parts of the cross spectrum using a combination of Lorentzian functions. This method assumes that the power and cross spectra of the source consist of multiple components that are coherent across different energy bands but incoherent with each other. 
    The constant phase-lag model proposed by \citet{2024MNRAS.527.9405M} assumes that the phase lags of individual Lorentzian components are constant with Fourier frequency. This approach simplifies the computation of phase lags by modeling the real and imaginary components of the cross spectrum using multiple Lorentzian functions.
    \footnote{The Real and Imaginary parts of a cross spectrum are expressed as: $\mathrm{Re}(\nu) = \sum_{i=1}^n C_i L(\nu; \nu_{0,i}, \Delta_i) \cos (2\pi k_i)$ and $\mathrm{Im}(\nu) = \sum_{i=1}^n C_i L(\nu;\nu_{0,i}, \Delta_i) \sin (2\pi k_i)$\,, where $L(\nu; \nu_{0,i}, \Delta_i)$ represents the Lorentzian functions with the centroid frequency $\nu_{0,i}$ and the FWHM $\Delta_i$, and $2\pi k_i$ are the phase lags for each component.}
    We refer readers to \citet{2024MNRAS.527.9405M} for the details of the method.
    
    %

     In the upper-right and lower-right panels of Fig.~\ref{Fig3}, we show the Real and Imaginary parts of the cross spectrum for Group 3 in the 4-10\,keV band, with the 2-4\,keV band as the reference. In the lower-left panel of Fig.~\ref{Fig3}, we show the phase lag versus Fourier frequency.
     Given that the power of the imaginary part of the cross spectrum is significantly smaller than that of the real part, we rotated the cross-vector by $45\degree$ to enhance the stability of the fitting process.
     We simultaneously fitted the Real and Imaginary parts of the cross spectrum using the same number of Lorentzian components as those employed in the PDS fitting, fixing the frequency and FWHM of each Lorentzian to the values derived from the PDS fits, and assuming the constant phase-lags model as described in \citet{2024MNRAS.527.9405M} and \citet{2025A&A...699A...9J}.
     Using this method, we obtained the phase lags associated with each variability component.

\subsection{Energy-dependent properties of the QPO and high-frequency humps}
\label{section 3.2}
    \begin{figure*}
    \centering
    \includegraphics[width=1\linewidth]{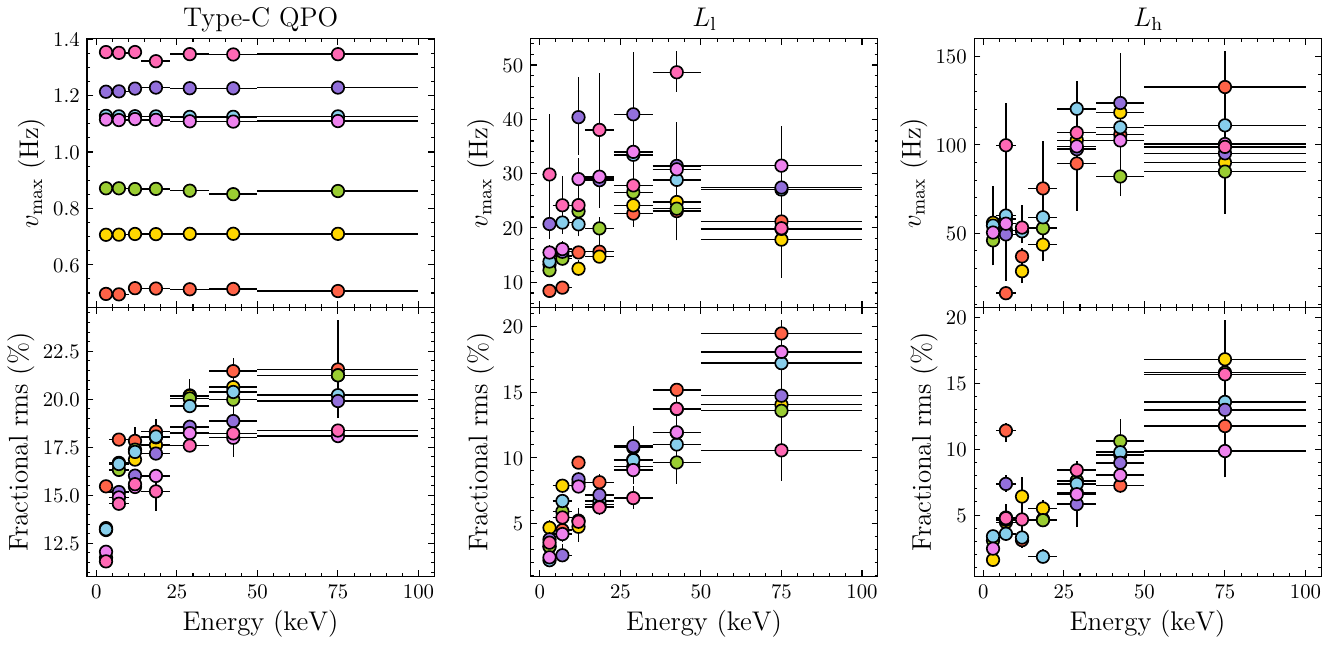}
        \caption{Energy dependence of the characteristic frequency and fractional rms of the type-C QPO and the two high-frequency humps ($L_{\rm l}$ and $L_\mathrm{h}$) for each data group. The color scheme for the 7 data groups follows that highlighted in Fig.~\ref{Fig2}.}
            \label{Fig4}%
    \end{figure*}

    To examine the energy-dependent properties of the variability components, we generated PDS for different energy bands: LE (2-4\,keV, 4-10\,keV), ME (10-14\,keV, 14-23\,keV), and HE (23-35\,keV, 35-50\,keV, 50-100\,keV). 
    Additionally, we computed the cross spectrum for each energy band, using the 2–4\,keV band as the reference.
    Following the previously described method, we obtained the characteristic frequency ($\nu_{\mathrm{max}}$\footnote{$\nu_{\mathrm{max}} = \sqrt{\nu_0^2 + (\sigma/2)^2}$\,, where $\nu_0$ and $\sigma$ represent the centroid frequency and FWHM of the Lorenzian function used to fit the component, respectively.}), fractional rms amplitude, and phase lag of type-C QPO, $L_{\rm l}$, and $L_{\rm h}$ in each energy band.

    In Fig.~\ref{Fig4}, we show the characteristic frequency ($\nu_{\rm max}$) and fractional rms amplitude of type-C QPO, $L_{\rm l}$, and $L_\mathrm{h}$ as a function of photon energy. 
    We observed that the characteristic frequency of type-C QPO remains constant across different energy bands, while its fractional rms amplitude exhibits a rapid increase with energy below 25\,keV, and then remains more or less constant at higher energies. This is consistent with the findings reported in \citet{2024ApJ...970L..33Y} and \citet{2024MNRAS.529.4624Y}.
    The behaviors of the two high-frequency humps, $L_{\rm l}$ and $L_\mathrm{h}$, exhibit notable similarities. Their characteristic frequencies increase with energy up to 25\,keV and then stabilize. Additionally, their fractional rms amplitudes consistently rise with energy towards higher energy bands, although a scatter is observed for $L_{\rm h}$ in the low-energy bands.

    \begin{figure}
    \centering
    \includegraphics[width=\hsize]{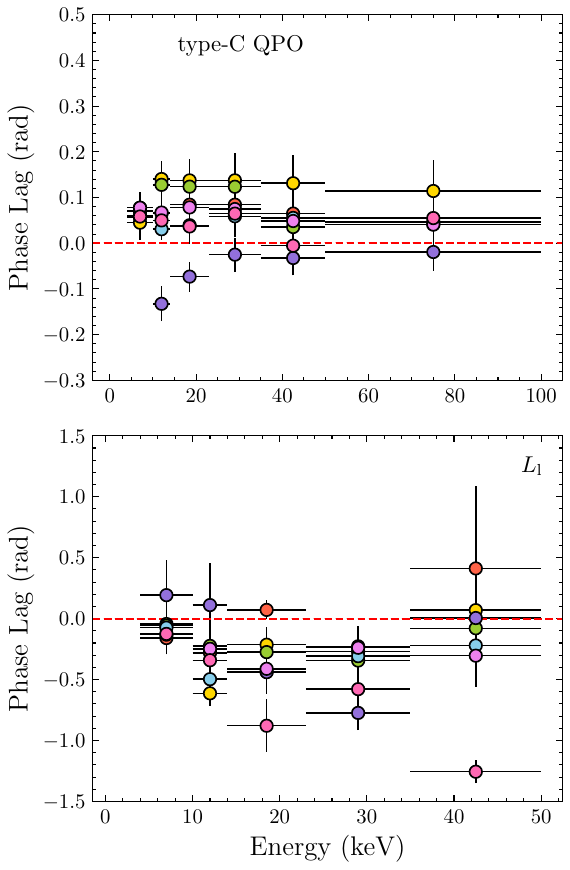}
        \caption{Energy dependence of the phase lag of the type-C QPOs and $L_{\rm l}$ for each data group. The color scheme for the 7 data groups follows that highlighted in Fig.~\ref{Fig2}.}
            \label{Fig5}
    \end{figure}

    In Fig.~\ref{Fig5}, we show the phase lags of type-C QPO and $L_{\rm l}$ as a function of photon energy. The phase lag of type-C QPO shows slight hard lags in most cases, remaining consistent across energy bands without significant variation. However, the lag spectrum of $L_{\rm l}$ is characterized primarily by a soft lag that varies with energy.
    %
    Owing to the limited signal-to-noise ratio, the phase lag associated with both $L_\mathrm{l}$ in 50-100\,keV range and $L_\mathrm{h}$ cannot be constrained well.

    \begin{figure}
    \centering
    \includegraphics[width=\hsize]{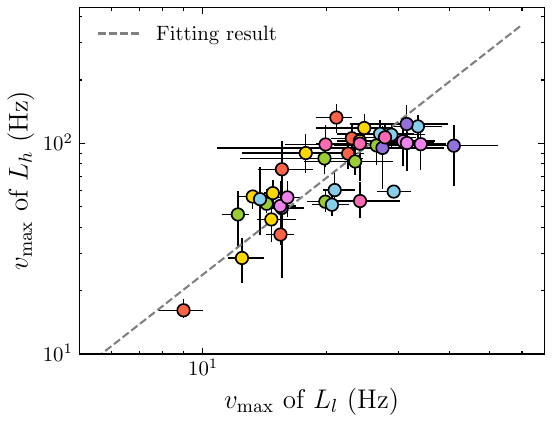}
        \caption{The relationship between the characteristic frequencies of the high-frequency humps,  $L_{\rm l}$ and $L_\mathrm{h}$. The dash line is the best-fitting result using an exponential function. }
            \label{Fig6}
    \end{figure}


\subsection{Measuring the mass and spin of the black hole with the Relativistic Precession Model}
\label{section 3.3}

    The RPM links three types of QPOs observed in BH-LMXBs to a combination of the fundamental frequencies of particle motion. Type-C QPOs are associated with the nodal precession frequency ($\nu_{\mathrm{nod}}$). The lower and upper HFQPOs corresponds to the periastron precession frequency ($\nu_{\mathrm{per}}$) and orbital frequency ($\nu_{\mathrm{\phi}}$), respectively \citep{1998ApJ...492L..59S, 1999PhRvL..82...17S, 1999ApJ...524L..63S, 2014MNRAS.437.2554M}.
    Assuming that these frequencies originate from the same radius ($r$), the orbital frequency, the periastron precession frequency, and the nodal precession frequency can be expressed as:

    \begin{equation}
        \nu_{\mathrm{\phi}} = \pm \frac{1}{2 \pi} \left (\frac{M}{r^3} \right) ^{1/2} \frac{1}{1 \pm a \left(\frac{M}{r}\right)^{3/2}} \, ,
            \label{equa 3}
    \end{equation}
    \begin{equation}
        \nu_{\mathrm{per}} = \nu_{\mathrm{\phi}} \left(1-\left(1-\frac{6 M}{r}-3 a^2\left(\frac{M}{r}\right)^2 \pm 8 a \left(\frac{M}{r}\right)^{3/2}\right)^{1/2}\right) \, ,
            \label{equa 4}
    \end{equation}
    \begin{equation}
        \nu_{\mathrm{nod}} = \nu_{\mathrm{\phi}} \left(1-\left(1+3a^2\left(\frac{M}{r}\right)^2 \mp 4a\left(\frac{M}{r}\right)^{3/2}\right)^{1/2}\right) \, ,
            \label{equa 5}
    \end{equation}
    where $M$ is the black hole mass and $a$ is the dimensionless spin parameter.
    If all three QPOs are detected simultaneously, we can determine the mass and spin of the black hole.
    Futhermore, \cite{2014MNRAS.444.2065I} found the analytical solution to the RPM system as follows:
    \begin{equation}
        r = \frac{2}{3}\frac{6-\Delta-5\Gamma+2\sqrt{2\left(\Delta-\Gamma\right)\left(3-\Delta-2\Gamma\right)}}{\left(\Delta+\Gamma-2\right)^2} \, ,
            \label{equa 6}
    \end{equation}
    \begin{equation}
        a = \pm \frac{r^{3/2}}{4}\left(\Delta+\Gamma-2+\frac{6}{r}\right) \, ,
            \label{equa 7}
    \end{equation}
    where $\Gamma$ and $\Delta$ are given by:
    \begin{equation}
        \Gamma = \left(1-\frac{\nu_{\mathrm{per}}}{\nu_{\mathrm{\phi}}}\right)^2 = 1-\frac{6}{r}\pm \frac{8a}{r^{3/2}}-\frac{3a^2}{r^2} \, ,
            \label{equa 8}
    \end{equation}
    \begin{equation}
        \Delta = \left(1-\frac{\nu_{\mathrm{nod}}}{\nu_{\mathrm{\phi}}}\right)^2 = 1\mp \frac{4a}{r^{3/2}}+\frac{3a^2}{r^2}\, .
            \label{equa 8}
    \end{equation}
    From this, the spin and mass can be determined from equations \ref{equa 7} and \ref{equa 3}.


    \citet{2014MNRAS.437.2554M} and \citet{2014MNRAS.439L..65M} investigated the HFQPOs and high-frequency humps of GRO J1655--40 and XTE J1550--564.
    The research findings indicate that the characteristic frequencies of the lower and upper high-frequency humps align with the periastron precession and orbital frequencies predicted by the RPM.
    \citet{2022MNRAS.514.2891Z} further proposed that the high-frequency humps and the HFQPOs may originate from the same variability component, with the coherence of this variability being influenced by the properties of the corona.
    Therefore, in principle, triplets consisting of two high-frequency humps and a type-C QPO can be used to constrain the mass and spin of the black hole.
    \citet{2021MNRAS.508.3104B} applied this method to the BH-LMXB MAXI J1820+070 and estimated its spin to be $0.799^{+0.016}_{-0.015}$.

    In \Swift, we detected two prominent high-frequency humps, $L_{\rm l}$ and $L_\mathrm{h}$, in each analyzed data group, accompanied by a type-C QPO. 
    \cite{2024MNRAS.529.4624Y} found that the frequency of $L_{\rm l}$ exhibits a strong correlation with the frequency of the type-C QPO, consistent with the PBK relation \citep{1999ApJ...520..262P}. Meanwhile, we found that the characteristic frequencies of the two high-frequency humps, $L_{\rm l}$ and $L_\mathrm{h}$, are also correlated, as shown in Fig.~\ref{Fig6}. Their relationship can be fitted with an exponential function as $\nu_h = (0.71 \pm 0.38) \nu_l ^{1.52 \pm 0.18}$.
    %
    %
    
\begin{figure}
    \centering
    \includegraphics[width=\hsize]{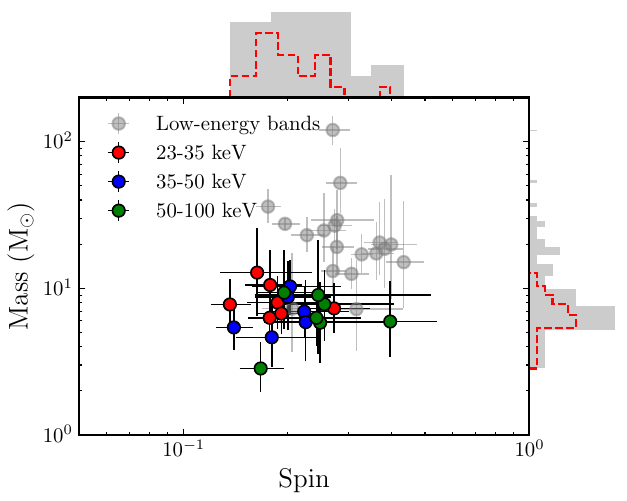}
        \caption{
        The mass and spin estimates derived from the relativistic precession model using the triplets composed of a type-C QPO and two high-frequency humps.The gray points represent the spin and mass values obtained using the triples from the energy bands below 23 keV. The distributions of the spin and mass can be seen at the top and to the right of the figure, the gray areas correspond to the distribution of the full-energy bands, while the red dashed lines represent the distribution considering only the high-energy bands above 23 keV.
        }
            \label{Fig7}
    \end{figure}

\begin{figure*}
    \centering
    \includegraphics[width=\hsize]{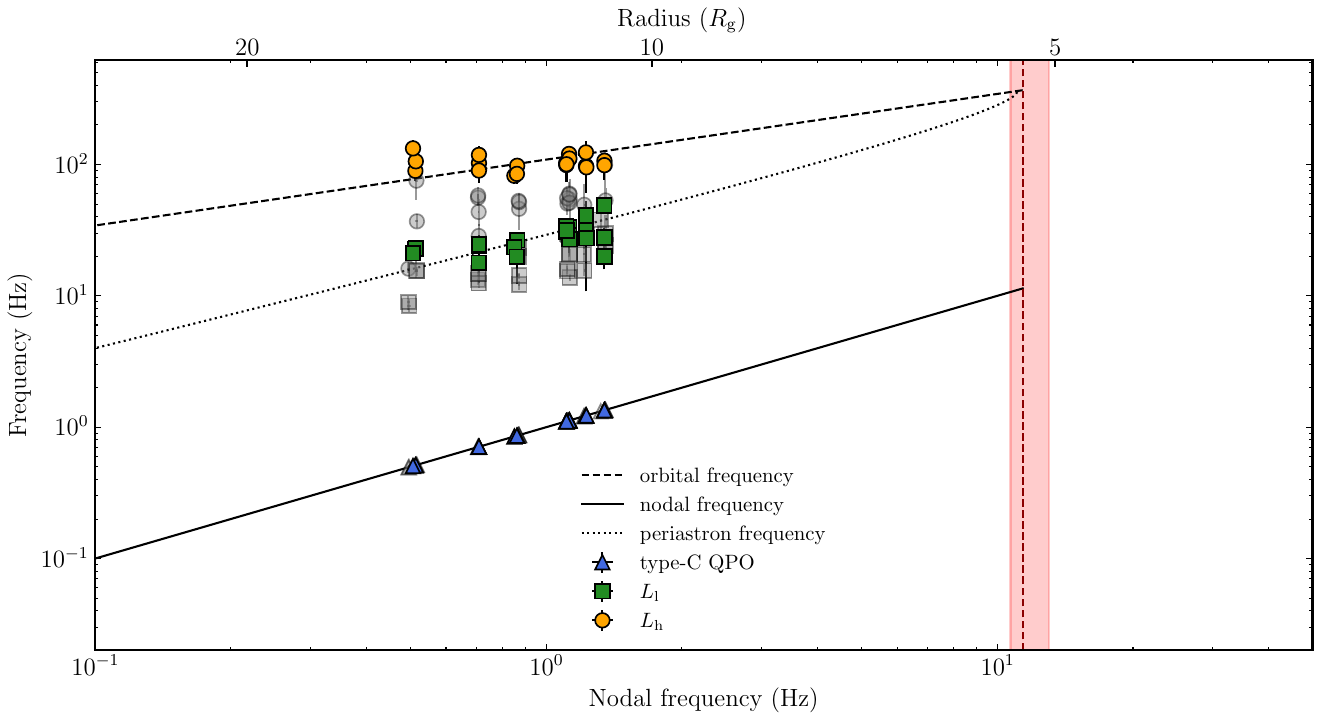}
        \caption{Nodal precession frequency (solid line), periastron precession frequency (dashed line), and orbital frequency (dotted line) as a function of the nodal precession frequency as predicted by the relativistic precession model. The lines are drawn for mass $ M = 7.12\,M_{\odot}$ and spin $a = 0.20$. The corresponding radii are given in the top $x$-axis. The blue triangles represent the characteristic frequencies of the type-C QPO of high-energy data. The green squares and yellow circles mark the characteristic frequencies of the two high-frequency humps, $L_{\rm l}$ and $L_{\rm h}$. Additionally, the gray triangles, squares, and circles denote the characteristic frequencies of the type-C QPO, $L_{\rm l}$, and $L_{\rm h}$ in the low-energy band, respectively. The vertical dashed line represents the nodal precession frequency at the ISCO, where the Keplerian frequency equals the periastron precession frequency and the red vertical band indicates its corresponding $1\sigma$ uncertainty.}
            \label{Fig8}
    \end{figure*}

    We then employed the RPM to estimate the mass and spin of the black hole in \Swift\ using triplets composed of two high-frequency humps and a type-C QPO observed in each data group and energy band.
    %
    %
    Following \cite{2014MNRAS.437.2554M}, we calculated the errors using a Monte Carlo: we simulated $10^5$ sets of three frequencies measured from the PDS of \Swift. We then solved the RPM equations for each set of the three frequencies and obtained distributions of mass, spin, and radius values. By fitting the distributions of these parameters, we derived measurements of the mass and spin.
    %
    %
    Fig.~\ref{Fig7} shows the mass and spin predicted by each of the triplets used in our analysis. The parameter distributions are represented as histograms located at the top and right of the figure. Using the full-energy band dataset, the mass and spin distributions are constrained within the ranges of $2.84 < M/M_{\odot} < 120.01$ and $0.14<a<0.43$, with median values of $10.30 \,M_{\odot}$ and $0.25$, respectively. When considering only the high-energy bands (greater than 25\,keV), where the characteristic frequencies of the two humps tend to stabilize, the mass and spin distributions are determined to be $2.84 < M/M_{\odot} < 13.98$ and $0.14<a<0.40$, with median values of $7.12 \,M_{\odot}$ and $0.20$, respectively. 
    Consequently, in Fig. \ref{Fig8}, we identify the Type-C QPO and the two high-frequency humps ($L_{\rm l}$ and $L_{\rm h}$) as the nodal precession frequency, periastron precession frequency, and orbital frequency, respectively. We also present the mass and spin measurements of $7.12 \, M_{\odot}$ and $0.20$, obtained via the RPM from the high-energy dataset, and plot these values as a function of the nodal precession frequency.
    It is apparent that most of the characteristic frequencies match well the frequencies predicted by the best-fitting RPM.

\section{Discussion}
\label{section 4}

    We have systematically investigated the energy-dependent characteristics of the two prominent high-frequency humps identified in the PDS of \Swift\ using \HXMT\ observations. Our analysis reveals that their characteristic frequencies display significant energy dependence, initially increasing with energy up to $\sim$30\,keV, followed by a plateau at higher energies. Their fractional rms amplitudes generally increase with energy.
    The phase lag associated with $L_{\rm l}$ typically exhibits a soft lag.
    Based on the assumption that the two high-frequency humps correspond to the periastron precession frequency and orbital frequency, with the type-C QPO being associated with the nodal precession frequency, we have derived estimates for both the black hole mass and spin parameters of \Swift. Below, we discuss our main results.

\subsection{Characteristics of the high-frequency humps}

    The energy-dependent fractional rms amplitude of the high-frequency hump was systematically investigated in GRS 1915+105 by \citet{2022MNRAS.514.2891Z}, using the RXTE observations in the energy band below 20\,keV. Their analysis revealed that the rms amplitude of the hump generally increases with energy.
    Thanks to the broad energy coverage and large effective area of \HXMT, we have extended this investigation to the high-frequency humps in \Swift\ up to 100\,keV.
    Both $L_{\rm l}$ and $L_\mathrm{h}$ exhibit an increase in the fractional rms amplitude with energy, eventually reaching approximately 15\% in the 50–100\,keV band. 
    The high rms values observed in the high energy bands suggest that the high-frequency humps originate from the corona, as neither the accretion disk nor the reflection component contributes significantly to the emission in these energy ranges  \citep{Gilfanov2010, 2024ApJ...970L..33Y}.
    %
    In both GX 339--4 \citep{2024MNRAS.527.5638Z} and GRS 1915+105 \citep{2022MNRAS.514.2891Z}, the rms amplitude of the hump is significantly stronger in the corona-dominated state, when the source exhibits a high corona temperature.
    Additionally, \cite{2003A&A...407.1039P} observed that the high-frequency hump in Cygnus X-1 nearly vanishes during the transition to the soft state.
    These results further indicate that the mechanism responsible for producing the hump is closely associated with the corona.
    We note that the energy-dependent fractional rms of the high-frequency hump closely resembles that of the HFQPO observed in BH-LMXBs, with the rms typically increasing with photon energy \citep{1997ApJ...482..993M, 2001ApJ...552L..49S, 2001ApJ...563..928M, 2013MNRAS.432...19B}.

    The phase lag associated with the high-frequency humps cannot be precisely constrained. We found that the phase lag of $L_{\rm l}$ is predominantly soft. Similar results have been observed for the phase lag of the lower kHz QPOs in Neutron Stars. Both the lower kHz QPOs in 4U 1608--52 \citep{1997ApJ...483L.115V, 1998ApJ...509L.145V} and 4U 1636--53 \citep{1999ApJ...514L..31K, 2020MNRAS.492.1399K} exhibit a soft phase lag.  
    %
     The soft lag may result from the photons emitted from the corona being Compton down-scattered by the cold plasma in the disk close to the corona \citep[e.g.,][]{2000ApJ...541..883R, 2007ApJ...661.1084F}.
    


\subsection{The Mass and Spin of \Swift}
\begin{table}
\centering
\caption[]{Black hole spin values measured from the Relativistic Precession Model. }
    \label{Table2}
    \resizebox{\linewidth}{!}{
    \begin{tabular}{ccc}
        \hline
        \noalign{\smallskip}
        Source & Spin  & Reference\\
        \noalign{\smallskip}
        \hline
        \noalign{\smallskip}
        GRO J1655--40 & $0.290 \pm 0.003$ & \cite{2014MNRAS.437.2554M} \\
        \noalign{\smallskip}
        XTE J1550--564 & $0.34 \pm 0.01$ & \cite{2014MNRAS.439L..65M} \\
        \noalign{\smallskip}
        XTE J1859+226 & $0.149 \pm 0.005$ & \cite{2022MNRAS.517.1469M} \\
        \noalign{\smallskip}
        MAXI J1820+070 $^a$ & $0.799_{-0.015}^{+0.016}$ & \cite{2021MNRAS.508.3104B} \\
        \noalign{\smallskip}
        H 1743--322 & $0.2 \sim 0.6$ & \cite{2018PAN....81..279T}\\
        \noalign{\smallskip}
        \Swift & $0.14 \sim 0.43$ &  this work \\
        \noalign{\smallskip}
        \hline
        \noalign{\smallskip}
    \end{tabular}}
    \footnotesize{
        \begin{minipage}{\linewidth}
            \raggedright
            Notes. $^a$ In \cite{2021MNRAS.508.3104B}, the authors used triplets composed of a low-frequency QPO and two broadband noise components to estimate the spin of the black hole.
        \end{minipage}
    }
\end{table}

    Previous studies have reported several measurements of the mass and spin of \Swift.
    \cite{2025A&A...693A.129M} conducted optical spectral observations of \Swift\ with GTC telescope to construct its mass function as $f(M_1)=2.77\pm 0.09 \, M_{\odot}$, establishing a lower mass limit of $3.12 \pm 0.10\, M_{\odot}$.
    \citet{2024ApJ...975..194D} analyzed the combined spectra from \textit{NICER} and \textit{NuSTAR} using the Two Component Advective Flow (TCAF) model, determining a mass of $10.2 \pm 0.4 \, M_{\odot}$ for the system. 
    The mass estimate obtained in our study is generally consistent with the ranges determined by \cite{2025A&A...693A.129M} and \cite{2024ApJ...975..194D}.

    In contrast to the high spin value of $a \sim 0.98$ obtained from reflection modeling \citep{2024arXiv240603834L}, our analysis revealed a relatively low spin value, specifically in the range of $ 0.14 \sim 0.43 $. A similar discrepancy is observed in GRO J1655-–40, where the spin value measured by \cite{2014MNRAS.437.2554M} using the RPM is $0.290 \pm 0.003$, while the spin estimates from X-ray spectral analysis are significantly higher, e.g., $a=0.65-0.75$ \citep{2006ApJ...636L.113S} and $a=0.94-0.98$ \citep{2009ApJ...697..900M}. 

    It is noteworthy that the black hole spins determined through X-ray reflection spectroscopy and thermal continuum fitting typically exhibit high values \citep[see][]{2021ARA&A..59..117R,2023ApJ...946...19D}, whereas those derived from the RPM tend to be lower (see Table \ref{Table2}). Both relativistic reflection \citep{1991MNRAS.249..352G, 2003AdSpR..32.2021Y, 2007ARA&A..45..441M} and thermal continuum fitting \citep{1997ApJ...482L.155Z, 2001MNRAS.325.1253G} require the assumption that the inner radius of the accretion disk coincides with ISCO of the black hole \citep{2009ApJ...707L..87T}. The black hole spin can be derived by determining the ISCO radius through measurements of the inner disk radius \citep{1972ApJ...178..347B, 1973blho.conf..343N}. However, the observed ISCO radius may be systematically biased towards smaller values due to radiation from the intra-ISCO region, i.e., plunging region (e.g. \citealp{2010ApJ...711..959N, 2010MNRAS.408..752P, 2024MNRAS.532.3395M}), resulting in an overestimation of the spin. Additionally, when using the reflection spectrum fitting method, neglecting the corona's scattering of the reflection component can result in an overestimation of the width of the iron line profile \citep{2017ApJ...836..119S}, subsequently leading to an inflated estimation of the spin value.
    %

    


    The RPM also relies on several key assumptions.
    Specifically, this model is based on the frequencies of test particles in the accretion disk around compact objects and does not account for hydrodynamical effects in the accretion flow that could affect those frequencies. The motion of test particles is unlikely to produce the broad humps observed in the PDS. 
    In this work, we have determined the mass and spin of \Swift\ to be $2.84 < M/M_{\odot} < 120.01$ and $0.14 < a < 0.43$, respectively, using the full-energy band dataset. Additionally, when considering only the high-energy bands, we obtained tighter constraints of $2.84 < M/M_{\odot} < 13.98$ and $0.14 < a < 0.40$ using the RPM.
    We find that the derived black hole mass and spin values vary across different energy bands, and the stabilization of characteristic frequencies in the high-energy regime serves as a viable reference for interpreting these parameters. As shown in Fig.~\ref{Fig4}, the characteristic frequencies of the high-frequency humps increase significantly with energy up to $\sim$25\,keV, above which they plateau. However, current research has not yet provided a satisfactory explanation for the stability of these characteristic frequencies within the context of particle precession.
    Moreover, the physical origin of QPOs in BH-LMXBs, and specifically whether they can be accurately described by the RPM, remain subjects of intense debate.
    %
    Despite these unresolved issues, we conclude that, compared to X-ray spectral fitting methods, the RPM offers an alternative and independent method for estimating black hole spin. Further systematic studies with this method are essential to enhance our understanding of its reliability and limitations.

\subsection{Disk truncation}
    The measurement of the inner disk radius from spectral analysis depends highly on the choice of spectral model, as different models are based on distinct geometric and physical assumptions. 
    \cite{2024arXiv240603834L} analyzed the \HXMT\ spectra of \Swift\ during the rising phase of the normal outburst using the model \texttt{Constant*tbabs*(diskbb+relxill+cutoffpl)}. They found that the inner disk radius shows no significant variation over time, with values about 3 $R_{\rm g}$.
    In contrast, \cite{2025arXiv250621131X} measured the inner disk radius of \Swift\ during the flare state with the model \texttt{Constant*tbabs*(thcomp*diskbb+relxillCp)}. They found that the radius gradually decreases from $\sim 50 \, R_{\rm g}$ to $\sim 7 \, R_{\rm g}$ as the QPO frequency increases from around 1\,Hz to 4.5\,Hz. When the QPO frequency exceeds 4.5\,Hz, the inner disk radius remains unchanged. 
    Using the RPM model, we infer the radius of ISCO to be $5.29^{+0.11}_{-0.23}\, R_g$. The characteristic radius associated with the QPOs is found to decrease from $\sim 15 \, R_{\rm g}$ to $\sim 11 \, R_{\rm g}$ as the spectrum softens (see Fig. \ref{Fig8}), suggesting a modest disk truncation.
    We fitted the NuSTAR\footnote{We exclude the \HXMT\ spectra due to the presence of instrument-related features near 5\,keV, which may interfere with the fitting of the reflection spectra.} spectrum (ObsID: 80902333004) using the model \texttt{Constant*tbabs(diskbb+relxill+cutoffpl)}, which was also employed by \cite{2024arXiv240603834L} for the analysis conducted prior to the flare state, with the spin parameter fixed at 0.20 (as measured by the RPM from the high-energy dataset). The spectral fitting results an inner disk radius ($R_{\rm in}$) of $1.7^{+1.1}_{-0.5}\, R_{\rm  ISCO}$, which is comparable to the radius of $2.18^{+0.05}_{-0.10} \, R_{\rm ISCO}$ measured from the RPM (detailed parameters are provided in Appendix C). 
    However, it should be noted that the RPM model is based on test particles and does not account for the structure of the disk.



\begin{acknowledgements}
      This work made use of data from the \HXMT\ mission, a project funded by China National Space Administration (CNSA) and the Chinese Academy of Sciences (CAS). This work is supported by the National Key R\&D Program of China (2021YFA0718500). 
      We acknowledge funding support from the National Natural Science Foundation of China under grants Nos.\ 12333007, 12122306, 12025301, \& 12103027, and the Strategic Priority Research Program of the Chinese Academy of Sciences.
      We acknowledge support from the China's  Space Origins Exploration Program.
\end{acknowledgements}

\bibliographystyle{aa}
\bibliography{cite}

\begin{thebibliography}{95}
\expandafter\ifx\csname natexlab\endcsname\relax\def\natexlab#1{#1}\fi

\bibitem[{{Abramowicz} \& {Klu{\'z}niak}(2001)}]{2001A&A...374L..19A}
{Abramowicz}, M.~A. \& {Klu{\'z}niak}, W. 2001, \aap, 374, L19

\bibitem[{{Alabarta} {et~al.}(2022){Alabarta}, {M{\'e}ndez}, {Garc{\'\i}a}, {Peirano}, {Altamirano}, {Zhang}, \& {Karpouzas}}]{2022MNRAS.514.2839A}
{Alabarta}, K., {M{\'e}ndez}, M., {Garc{\'\i}a}, F., {et~al.} 2022, \mnras, 514, 2839

\bibitem[{{Bardeen} {et~al.}(1972){Bardeen}, {Press}, \& {Teukolsky}}]{1972ApJ...178..347B}
{Bardeen}, J.~M., {Press}, W.~H., \& {Teukolsky}, S.~A. 1972, \apj, 178, 347

\bibitem[{{Belloni} \& {Hasinger}(1990)}]{Belloni1990}
{Belloni}, T. \& {Hasinger}, G. 1990, \aap, 230, 103

\bibitem[{{Belloni} {et~al.}(2002){Belloni}, {Psaltis}, \& {van der Klis}}]{2002ApJ...572..392B}
{Belloni}, T., {Psaltis}, D., \& {van der Klis}, M. 2002, \apj, 572, 392

\bibitem[{{Belloni} \& {Altamirano}(2013)}]{2013MNRAS.432...19B}
{Belloni}, T.~M. \& {Altamirano}, D. 2013, \mnras, 432, 19

\bibitem[{{Belloni} \& {Motta}(2016)}]{Belloni2016}
{Belloni}, T.~M. \& {Motta}, S.~E. 2016, in Astrophysics and Space Science Library, Vol. 440, Astrophysics of Black Holes: From Fundamental Aspects to Latest Developments, ed. C.~{Bambi}, 61

\bibitem[{{Belloni} {et~al.}(2012){Belloni}, {Sanna}, \& {M{\'e}ndez}}]{2012MNRAS.426.1701B}
{Belloni}, T.~M., {Sanna}, A., \& {M{\'e}ndez}, M. 2012, \mnras, 426, 1701

\bibitem[{{Bhargava} {et~al.}(2021){Bhargava}, {Belloni}, {Bhattacharya}, {Motta}, \& {Ponti.}}]{2021MNRAS.508.3104B}
{Bhargava}, Y., {Belloni}, T., {Bhattacharya}, D., {Motta}, S., \& {Ponti.}, G. 2021, \mnras, 508, 3104

\bibitem[{{Burridge} {et~al.}(2025){Burridge}, {Miller-Jones}, {Bahramian}, {Prabu}, {Streeter}, {Castro Segura}, {Corral Santana}, {Knigge}, {Tremou}, {Carotenuto}, {Fender}, \& {Saikia}}]{2025arXiv250206448B}
{Burridge}, B.~J., {Miller-Jones}, J. C.~A., {Bahramian}, A., {et~al.} 2025, arXiv e-prints, arXiv:2502.06448

\bibitem[{{Cao} {et~al.}(2020){Cao}, {Jiang}, {Meng}, {Zhang}, {Luo}, {Yang}, {Zhang}, {Gu}, {Sun}, {Liu}, {Yang}, {Li}, {Tan}, {Liu}, {Du}, {Lu}, {Xu}, {Guan}, {Zhang}, {Wang}, {Li}, {Zhang}, {Wen}, {Qu}, {Song}, {Li}, {Ge}, {Zhou}, {Xiong}, {Zhang}, {Zhang}, {Cheng}, {Zhang}, {Li}, {Liang}, {Gao}, {Yang}, {Liu}, {Liu}, {Yang}, \& {Zhang}}]{2020SCPMA..6349504C}
{Cao}, X., {Jiang}, W., {Meng}, B., {et~al.} 2020, Science China Physics, Mechanics, and Astronomy, 63, 249504

\bibitem[{{Casella} {et~al.}(2005){Casella}, {Belloni}, \& {Stella}}]{2005ApJ...629..403C}
{Casella}, P., {Belloni}, T., \& {Stella}, L. 2005, \apj, 629, 403

\bibitem[{{Chen} {et~al.}(2020){Chen}, {Cui}, {Li}, {Wang}, {Xu}, {Lu}, {Wang}, {Chen}, {Han}, {Hu}, {Zhang}, {Huo}, {Yang}, {Li}, {Lu}, {Zhang}, {Li}, {Zhang}, {Xiong}, {Zhang}, {Xue}, {Zhao}, {Zhu}, {Zhu}, {Liu}, {Yang}, \& {Zhang}}]{2020SCPMA..6349505C}
{Chen}, Y., {Cui}, W., {Li}, W., {et~al.} 2020, Science China Physics, Mechanics, and Astronomy, 63, 249505

\bibitem[{{Debnath} {et~al.}(2024){Debnath}, {Nath}, {Chatterjee}, {Chatterjee}, \& {Chang}}]{2024ApJ...975..194D}
{Debnath}, D., {Nath}, S.~K., {Chatterjee}, D., {Chatterjee}, K., \& {Chang}, H.-K. 2024, \apj, 975, 194

\bibitem[{{Done} {et~al.}(2007){Done}, {Gierli{\'n}ski}, \& {Kubota}}]{2007A&ARv..15....1D}
{Done}, C., {Gierli{\'n}ski}, M., \& {Kubota}, A. 2007, \aapr, 15, 1

\bibitem[{{Draghis} {et~al.}(2023){Draghis}, {Miller}, {Zoghbi}, {Reynolds}, {Costantini}, {Gallo}, \& {Tomsick}}]{2023ApJ...946...19D}
{Draghis}, P.~A., {Miller}, J.~M., {Zoghbi}, A., {et~al.} 2023, \apj, 946, 19

\bibitem[{{du Buisson} {et~al.}(2019){du Buisson}, {Motta}, \& {Fender}}]{2019MNRAS.486.4485D}
{du Buisson}, L., {Motta}, S., \& {Fender}, R. 2019, \mnras, 486, 4485

\bibitem[{{Falanga} \& {Titarchuk}(2007)}]{2007ApJ...661.1084F}
{Falanga}, M. \& {Titarchuk}, L. 2007, \apj, 661, 1084

\bibitem[{{Fogantini} {et~al.}(2025){Fogantini}, {Garc{\'\i}a}, {M{\'e}ndez}, {K{\"o}nig}, \& {Wilms}}]{2025arXiv250303078F}
{Fogantini}, F.~A., {Garc{\'\i}a}, F., {M{\'e}ndez}, M., {K{\"o}nig}, O., \& {Wilms}, J. 2025, arXiv e-prints, arXiv:2503.03078

\bibitem[{{George} \& {Fabian}(1991)}]{1991MNRAS.249..352G}
{George}, I.~M. \& {Fabian}, A.~C. 1991, \mnras, 249, 352

\bibitem[{{Gierli{\'n}ski} {et~al.}(2001){Gierli{\'n}ski}, {Macio{\l}ek-Nied{\'z}wiecki}, \& {Ebisawa}}]{2001MNRAS.325.1253G}
{Gierli{\'n}ski}, M., {Macio{\l}ek-Nied{\'z}wiecki}, A., \& {Ebisawa}, K. 2001, \mnras, 325, 1253

\bibitem[{Gilfanov(2010)}]{Gilfanov2010}
Gilfanov, M. 2010, X-Ray Emission from Black-Hole Binaries, ed. T.~Belloni (Berlin, Heidelberg: Springer Berlin Heidelberg), 17--51

\bibitem[{{Homan} {et~al.}(2005){Homan}, {Miller}, {Wijnands}, {van der Klis}, {Belloni}, {Steeghs}, \& {Lewin}}]{2005ApJ...623..383H}
{Homan}, J., {Miller}, J.~M., {Wijnands}, R., {et~al.} 2005, \apj, 623, 383

\bibitem[{{Homan} {et~al.}(2001){Homan}, {Wijnands}, {van der Klis}, {Belloni}, {van Paradijs}, {Klein-Wolt}, {Fender}, \& {M{\'e}ndez}}]{Homan2001}
{Homan}, J., {Wijnands}, R., {van der Klis}, M., {et~al.} 2001, \apjs, 132, 377

\bibitem[{{Ingram} {et~al.}(2009){Ingram}, {Done}, \& {Fragile}}]{2009MNRAS.397L.101I}
{Ingram}, A., {Done}, C., \& {Fragile}, P.~C. 2009, \mnras, 397, L101

\bibitem[{{Ingram} \& {Motta}(2014)}]{2014MNRAS.444.2065I}
{Ingram}, A. \& {Motta}, S. 2014, \mnras, 444, 2065

\bibitem[{{Ingram} {et~al.}(2016){Ingram}, {van der Klis}, {Middleton}, {Done}, {Altamirano}, {Heil}, {Uttley}, \& {Axelsson}}]{2016MNRAS.461.1967I}
{Ingram}, A., {van der Klis}, M., {Middleton}, M., {et~al.} 2016, \mnras, 461, 1967

\bibitem[{{Ingram} \& {Motta}(2019)}]{2019NewAR..8501524I}
{Ingram}, A.~R. \& {Motta}, S.~E. 2019, \nar, 85, 101524

\bibitem[{{Jin} {et~al.}(2025){Jin}, {M{\'e}ndez}, {Garc{\'\i}a}, {Altamirano}, {Zhang}, \& {Rout}}]{2025A&A...699A...9J}
{Jin}, P., {M{\'e}ndez}, M., {Garc{\'\i}a}, F., {et~al.} 2025, \aap, 699, A9

\bibitem[{{Kaaret} {et~al.}(1999){Kaaret}, {Piraino}, {Ford}, \& {Santangelo}}]{1999ApJ...514L..31K}
{Kaaret}, P., {Piraino}, S., {Ford}, E.~C., \& {Santangelo}, A. 1999, \apjl, 514, L31

\bibitem[{{Kalemci} {et~al.}(2003){Kalemci}, {Tomsick}, {Rothschild}, {Pottschmidt}, {Corbel}, {Wijnands}, {Miller}, \& {Kaaret}}]{2003ApJ...586..419K}
{Kalemci}, E., {Tomsick}, J.~A., {Rothschild}, R.~E., {et~al.} 2003, \apj, 586, 419

\bibitem[{{Karpouzas} {et~al.}(2020){Karpouzas}, {M{\'e}ndez}, {Ribeiro}, {Altamirano}, {Blaes}, \& {Garc{\'\i}a}}]{2020MNRAS.492.1399K}
{Karpouzas}, K., {M{\'e}ndez}, M., {Ribeiro}, E.~M., {et~al.} 2020, \mnras, 492, 1399

\bibitem[{{Kato}(1990)}]{1990PASJ...42...99K}
{Kato}, S. 1990, \pasj, 42, 99

\bibitem[{{Kennea} \& {Swift Team}(2023)}]{2023GCN.34540....1K}
{Kennea}, J.~A. \& {Swift Team}. 2023, GRB Coordinates Network, 34540, 1

\bibitem[{{Liu} {et~al.}(2020){Liu}, {Zhang}, {Li}, {Lu}, {Chang}, {Li}, {Zhang}, {Jin}, {Yu}, {Zhang}, {Fu}, {Chen}, {Ji}, {Xu}, {Deng}, {Shang}, {Liu}, {Lu}, {Zhang}, {Dong}, {Li}, {Wu}, {Li}, {Wang}, {Wu}, {Zhang}, {Zhang}, {Xiong}, {Liu}, {Zhang}, {Liu}, {Yang}, \& {Zhang}}]{2020SCPMA..6349503L}
{Liu}, C., {Zhang}, Y., {Li}, X., {et~al.} 2020, Science China Physics, Mechanics, and Astronomy, 63, 249503

\bibitem[{{Liu} {et~al.}(2024){Liu}, {Xu}, {Zhang}, {Yu}, {Huang}, {Tao}, {Zhang}, {Yang}, {Zhao}, {Qu}, \& {Song}}]{2024arXiv240603834L}
{Liu}, H.-X., {Xu}, Y.-J., {Zhang}, S.-N., {et~al.} 2024, arXiv e-prints, arXiv:2406.03834

\bibitem[{{Ma} {et~al.}(2021){Ma}, {Tao}, {Zhang}, {Zhang}, {Bu}, {Ge}, {Chen}, {Qu}, {Zhang}, {Lu}, {Song}, {Yang}, {Yuan}, {Cai}, {Cao}, {Chang}, {Chen}, {Chen}, {Chen}, {Chen}, {Chen}, {Cui}, {Cui}, {Deng}, {Dong}, {Du}, {Fu}, {Gao}, {Gao}, {Gao}, {Gu}, {Guan}, {Guo}, {Han}, {Huang}, {Huo}, {Ji}, {Jia}, {Jiang}, {Jiang}, {Jin}, {Jin}, {Kong}, {Li}, {Li}, {Li}, {Li}, {Li}, {Li}, {Li}, {Li}, {Li}, {Li}, {Li}, {Liang}, {Liao}, {Liu}, {Liu}, {Liu}, {Liu}, {Liu}, {Liu}, {Lu}, {Lu}, {Luo}, {Luo}, {Meng}, {Nang}, {Nie}, {Ou}, {Sai}, {Shang}, {Song}, {Sun}, {Tan}, {Tuo}, {Wang}, {Wang}, {Wang}, {Wang}, {Wang}, {Wang}, {Wen}, {Wu}, {Wu}, {Wu}, {Xiao}, {Xiao}, {Xie}, {Xiong}, {Xu}, {Xu}, {Yang}, {Yang}, {Yang}, {Yi}, {Yin}, {You}, {Zhang}, {Zhang}, {Zhang}, {Zhang}, {Zhang}, {Zhang}, {Zhang}, {Zhang}, {Zhang}, {Zhang}, {Zhang}, {Zhang}, {Zhang}, {Zhang}, {Zhang}, {Zhang}, {Zhao}, {Zhao}, {Zheng}, {Zhou}, {Zhou}, {Zhu}, {Zhu}, \& {Zhuang}}]{2021NatAs...5...94M}
{Ma}, X., {Tao}, L., {Zhang}, S.-N., {et~al.} 2021, Nature Astronomy, 5, 94

\bibitem[{{Marcel} \& {Neilsen}(2021)}]{2021ApJ...906..106M}
{Marcel}, G. \& {Neilsen}, J. 2021, \apj, 906, 106

\bibitem[{{Mata S{\'a}nchez} {et~al.}(2024){Mata S{\'a}nchez}, {Mu{\~n}oz-Darias}, {Armas Padilla}, {Casares}, \& {Torres}}]{2024A&A...682L...1M}
{Mata S{\'a}nchez}, D., {Mu{\~n}oz-Darias}, T., {Armas Padilla}, M., {Casares}, J., \& {Torres}, M.~A.~P. 2024, \aap, 682, L1

\bibitem[{{Mata S{\'a}nchez} {et~al.}(2025){Mata S{\'a}nchez}, {Torres}, {Casares}, {Mu{\~n}oz-Darias}, {Armas Padilla}, \& {Yanes-Rizo}}]{2025A&A...693A.129M}
{Mata S{\'a}nchez}, D., {Torres}, M.~A.~P., {Casares}, J., {et~al.} 2025, \aap, 693, A129

\bibitem[{{M{\'e}ndez} {et~al.}(2013){M{\'e}ndez}, {Altamirano}, {Belloni}, \& {Sanna}}]{2013MNRAS.435.2132M}
{M{\'e}ndez}, M., {Altamirano}, D., {Belloni}, T., \& {Sanna}, A. 2013, \mnras, 435, 2132

\bibitem[{{M{\'e}ndez} {et~al.}(2022){M{\'e}ndez}, {Karpouzas}, {Garc{\'\i}a}, {Zhang}, {Zhang}, {Belloni}, \& {Altamirano}}]{2022NatAs...6..577M}
{M{\'e}ndez}, M., {Karpouzas}, K., {Garc{\'\i}a}, F., {et~al.} 2022, Nature Astronomy, 6, 577

\bibitem[{{M{\'e}ndez} {et~al.}(2024){M{\'e}ndez}, {Peirano}, {Garc{\'\i}a}, {Belloni}, {Altamirano}, \& {Alabarta}}]{2024MNRAS.527.9405M}
{M{\'e}ndez}, M., {Peirano}, V., {Garc{\'\i}a}, F., {et~al.} 2024, \mnras, 527, 9405

\bibitem[{{Miller}(2007)}]{2007ARA&A..45..441M}
{Miller}, J.~M. 2007, \araa, 45, 441

\bibitem[{{Miller} {et~al.}(2009){Miller}, {Reynolds}, {Fabian}, {Miniutti}, \& {Gallo}}]{2009ApJ...697..900M}
{Miller}, J.~M., {Reynolds}, C.~S., {Fabian}, A.~C., {Miniutti}, G., \& {Gallo}, L.~C. 2009, \apj, 697, 900

\bibitem[{{Miller} {et~al.}(2001){Miller}, {Wijnands}, {Homan}, {Belloni}, {Pooley}, {Corbel}, {Kouveliotou}, {van der Klis}, \& {Lewin}}]{2001ApJ...563..928M}
{Miller}, J.~M., {Wijnands}, R., {Homan}, J., {et~al.} 2001, \apj, 563, 928

\bibitem[{{Molteni} {et~al.}(1996){Molteni}, {Sponholz}, \& {Chakrabarti}}]{1996ApJ...457..805M}
{Molteni}, D., {Sponholz}, H., \& {Chakrabarti}, S.~K. 1996, \apj, 457, 805

\bibitem[{{Morgan} {et~al.}(1997){Morgan}, {Remillard}, \& {Greiner}}]{1997ApJ...482..993M}
{Morgan}, E.~H., {Remillard}, R.~A., \& {Greiner}, J. 1997, \apj, 482, 993

\bibitem[{{Motta} {et~al.}(2011){Motta}, {Mu{\~n}oz-Darias}, {Casella}, {Belloni}, \& {Homan}}]{Motta2011}
{Motta}, S., {Mu{\~n}oz-Darias}, T., {Casella}, P., {Belloni}, T., \& {Homan}, J. 2011, \mnras, 418, 2292

\bibitem[{{Motta} {et~al.}(2022){Motta}, {Belloni}, {Stella}, {Pappas}, {Casares}, {Mu{\~n}oz-Darias}, {Torres}, \& {Yanes-Rizo}}]{2022MNRAS.517.1469M}
{Motta}, S.~E., {Belloni}, T., {Stella}, L., {et~al.} 2022, \mnras, 517, 1469

\bibitem[{{Motta} {et~al.}(2014{\natexlab{a}}){Motta}, {Belloni}, {Stella}, {Mu{\~n}oz-Darias}, \& {Fender}}]{2014MNRAS.437.2554M}
{Motta}, S.~E., {Belloni}, T.~M., {Stella}, L., {Mu{\~n}oz-Darias}, T., \& {Fender}, R. 2014{\natexlab{a}}, \mnras, 437, 2554

\bibitem[{{Motta} {et~al.}(2015){Motta}, {Casella}, {Henze}, {Mu{\~n}oz-Darias}, {Sanna}, {Fender}, \& {Belloni}}]{2015MNRAS.447.2059M}
{Motta}, S.~E., {Casella}, P., {Henze}, M., {et~al.} 2015, \mnras, 447, 2059

\bibitem[{{Motta} {et~al.}(2014{\natexlab{b}}){Motta}, {Munoz-Darias}, {Sanna}, {Fender}, {Belloni}, \& {Stella}}]{2014MNRAS.439L..65M}
{Motta}, S.~E., {Munoz-Darias}, T., {Sanna}, A., {et~al.} 2014{\natexlab{b}}, \mnras, 439, L65

\bibitem[{{Mummery} \& {Stone}(2024)}]{2024MNRAS.532.3395M}
{Mummery}, A. \& {Stone}, J.~M. 2024, \mnras, 532, 3395

\bibitem[{{Negoro} {et~al.}(2023){Negoro}, {Serino}, {Nakajima}, {Kobayashi}, {Tanaka}, {Soejima}, {Kudo}, {Mihara}, {Kawamuro}, {Yamada}, {Tamagawa}, {Kawai}, {Matsuoka}, {Sakamoto}, {Sugita}, {Hiramatsu}, {Nishikawa}, {Yoshida}, {Tsuboi}, {Urabe}, {Nawa}, {Nemoto}, {Shidatsu}, {Takahashi}, {Niwano}, {Sato}, {Higuchi}, {Yatsu}, {Nakahira}, {Ueno}, {Tomida}, {Ishikawa}, {Ogawa}, {Kurihara}, {Ueda}, {Setoguchi}, {Yoshitake}, {Nakatani}, {Yamauchi}, {Hagiwara}, {Umeki}, {Otsuki}, {Yamaoka}, {Kawakubo}, {Sugizaki}, {Iwakiri}, \& {MAXI Team}}]{2023GCN.34544....1N}
{Negoro}, H., {Serino}, M., {Nakajima}, M., {et~al.} 2023, GRB Coordinates Network, 34544, 1

\bibitem[{{Noble} {et~al.}(2010){Noble}, {Krolik}, \& {Hawley}}]{2010ApJ...711..959N}
{Noble}, S.~C., {Krolik}, J.~H., \& {Hawley}, J.~F. 2010, \apj, 711, 959

\bibitem[{{Novikov} \& {Thorne}(1973)}]{1973blho.conf..343N}
{Novikov}, I.~D. \& {Thorne}, K.~S. 1973, in Black Holes (Les Astres Occlus), ed. C.~{Dewitt} \& B.~S. {Dewitt}, 343--450

\bibitem[{{Nowak}(2000)}]{2000MNRAS.318..361N}
{Nowak}, M.~A. 2000, \mnras, 318, 361

\bibitem[{{Penna} {et~al.}(2010){Penna}, {McKinney}, {Narayan}, {Tchekhovskoy}, {Shafee}, \& {McClintock}}]{2010MNRAS.408..752P}
{Penna}, R.~F., {McKinney}, J.~C., {Narayan}, R., {et~al.} 2010, \mnras, 408, 752

\bibitem[{{Pottschmidt} {et~al.}(2003){Pottschmidt}, {Wilms}, {Nowak}, {Pooley}, {Gleissner}, {Heindl}, {Smith}, {Remillard}, \& {Staubert}}]{2003A&A...407.1039P}
{Pottschmidt}, K., {Wilms}, J., {Nowak}, M.~A., {et~al.} 2003, \aap, 407, 1039

\bibitem[{{Psaltis} {et~al.}(1999){Psaltis}, {Belloni}, \& {van der Klis}}]{1999ApJ...520..262P}
{Psaltis}, D., {Belloni}, T., \& {van der Klis}, M. 1999, \apj, 520, 262

\bibitem[{{Reig} {et~al.}(2000){Reig}, {Belloni}, {van der Klis}, {M{\'e}ndez}, {Kylafis}, \& {Ford}}]{2000ApJ...541..883R}
{Reig}, P., {Belloni}, T., {van der Klis}, M., {et~al.} 2000, \apj, 541, 883

\bibitem[{{Remillard} \& {McClintock}(2006)}]{2006ARA&A..44...49R}
{Remillard}, R.~A. \& {McClintock}, J.~E. 2006, \araa, 44, 49

\bibitem[{{Remillard} {et~al.}(2006){Remillard}, {McClintock}, {Orosz}, \& {Levine}}]{2006ApJ...637.1002R}
{Remillard}, R.~A., {McClintock}, J.~E., {Orosz}, J.~A., \& {Levine}, A.~M. 2006, \apj, 637, 1002

\bibitem[{{Remillard} {et~al.}(1999){Remillard}, {Morgan}, {McClintock}, {Bailyn}, \& {Orosz}}]{1999ApJ...522..397R}
{Remillard}, R.~A., {Morgan}, E.~H., {McClintock}, J.~E., {Bailyn}, C.~D., \& {Orosz}, J.~A. 1999, \apj, 522, 397

\bibitem[{{Remillard} {et~al.}(2002{\natexlab{a}}){Remillard}, {Muno}, {McClintock}, \& {Orosz}}]{2002ApJ...580.1030R}
{Remillard}, R.~A., {Muno}, M.~P., {McClintock}, J.~E., \& {Orosz}, J.~A. 2002{\natexlab{a}}, \apj, 580, 1030

\bibitem[{{Remillard} {et~al.}(2002{\natexlab{b}}){Remillard}, {Sobczak}, {Muno}, \& {McClintock}}]{2002ApJ...564..962R}
{Remillard}, R.~A., {Sobczak}, G.~J., {Muno}, M.~P., \& {McClintock}, J.~E. 2002{\natexlab{b}}, \apj, 564, 962

\bibitem[{{Reynolds}(2021)}]{2021ARA&A..59..117R}
{Reynolds}, C.~S. 2021, \araa, 59, 117

\bibitem[{{Samimi} {et~al.}(1979){Samimi}, {Share}, {Wood}, {Yentis}, {Meekins}, {Evans}, {Shulman}, {Byram}, {Chubb}, \& {Friedman}}]{1979Natur.278..434S}
{Samimi}, J., {Share}, G.~H., {Wood}, K., {et~al.} 1979, \nat, 278, 434

\bibitem[{{Shafee} {et~al.}(2006){Shafee}, {McClintock}, {Narayan}, {Davis}, {Li}, \& {Remillard}}]{2006ApJ...636L.113S}
{Shafee}, R., {McClintock}, J.~E., {Narayan}, R., {et~al.} 2006, \apjl, 636, L113

\bibitem[{{Steiner} {et~al.}(2017){Steiner}, {Garc{\'\i}a}, {Eikmann}, {McClintock}, {Brenneman}, {Dauser}, \& {Fabian}}]{2017ApJ...836..119S}
{Steiner}, J.~F., {Garc{\'\i}a}, J.~A., {Eikmann}, W., {et~al.} 2017, \apj, 836, 119

\bibitem[{{Stella} \& {Vietri}(1998)}]{1998ApJ...492L..59S}
{Stella}, L. \& {Vietri}, M. 1998, \apjl, 492, L59

\bibitem[{{Stella} \& {Vietri}(1999)}]{1999PhRvL..82...17S}
{Stella}, L. \& {Vietri}, M. 1999, \prl, 82, 17

\bibitem[{{Stella} {et~al.}(1999){Stella}, {Vietri}, \& {Morsink}}]{1999ApJ...524L..63S}
{Stella}, L., {Vietri}, M., \& {Morsink}, S.~M. 1999, \apjl, 524, L63

\bibitem[{{Stevens} \& {Uttley}(2016)}]{2016MNRAS.460.2796S}
{Stevens}, A.~L. \& {Uttley}, P. 2016, \mnras, 460, 2796

\bibitem[{{Strohmayer}(2001)}]{2001ApJ...552L..49S}
{Strohmayer}, T.~E. 2001, \apjl, 552, L49

\bibitem[{{Tagger} \& {Pellat}(1999)}]{1999A&A...349.1003T}
{Tagger}, M. \& {Pellat}, R. 1999, \aap, 349, 1003

\bibitem[{{Tomsick} {et~al.}(2009){Tomsick}, {Yamaoka}, {Corbel}, {Kaaret}, {Kalemci}, \& {Migliari}}]{2009ApJ...707L..87T}
{Tomsick}, J.~A., {Yamaoka}, K., {Corbel}, S., {et~al.} 2009, \apjl, 707, L87

\bibitem[{{Trudolyubov}(2001)}]{2001ApJ...558..276T}
{Trudolyubov}, S.~P. 2001, \apj, 558, 276

\bibitem[{{Tursunov} \& {Kolo{\v{s}}}(2018)}]{2018PAN....81..279T}
{Tursunov}, A.~A. \& {Kolo{\v{s}}}, M. 2018, Physics of Atomic Nuclei, 81, 279

\bibitem[{{van der Klis} {et~al.}(1987){van der Klis}, {Hasinger}, {Stella}, {Langmeier}, {van Paradijs}, \& {Lewin}}]{1987ApJ...319L..13V}
{van der Klis}, M., {Hasinger}, G., {Stella}, L., {et~al.} 1987, \apjl, 319, L13

\bibitem[{{van der Klis} {et~al.}(1985){van der Klis}, {Jansen}, {van Paradijs}, {Lewin}, {van den Heuvel}, {Trumper}, \& {Szatjno}}]{1985Natur.316..225V}
{van der Klis}, M., {Jansen}, F., {van Paradijs}, J., {et~al.} 1985, \nat, 316, 225

\bibitem[{{Vaughan} {et~al.}(1997){Vaughan}, {van der Klis}, {M{\'e}ndez}, {van Paradijs}, {Wijnands}, {Lewin}, {Lamb}, {Psaltis}, {Kuulkers}, \& {Oosterbroek}}]{1997ApJ...483L.115V}
{Vaughan}, B.~A., {van der Klis}, M., {M{\'e}ndez}, M., {et~al.} 1997, \apjl, 483, L115

\bibitem[{{Vaughan} {et~al.}(1998){Vaughan}, {van der Klis}, {M{\'e}ndez}, {van Paradijs}, {Wijnands}, {Lewin}, {Lamb}, {Psaltis}, {Kuulkers}, \& {Oosterbroek}}]{1998ApJ...509L.145V}
{Vaughan}, B.~A., {van der Klis}, M., {M{\'e}ndez}, M., {et~al.} 1998, \apjl, 509, L145

\bibitem[{{Vikhlinin} {et~al.}(1995){Vikhlinin}, {Churazov}, {Gilfanov}, {Sunyaev}, {Finoguenov}, {Dyachkov}, {Kremnev}, {Sukhanov}, {Ballet}, {Goldwurm}, {Cordier}, {Claret}, {Denis}, {Olive}, {Roques}, \& {Mandrou}}]{1995ApJ...441..779V}
{Vikhlinin}, A., {Churazov}, E., {Gilfanov}, M., {et~al.} 1995, \apj, 441, 779

\bibitem[{{Wijnands} {et~al.}(1999){Wijnands}, {Homan}, \& {van der Klis}}]{1999ApJ...526L..33W}
{Wijnands}, R., {Homan}, J., \& {van der Klis}, M. 1999, \apjl, 526, L33

\bibitem[{{Wood} {et~al.}(2024){Wood}, {Miller-Jones}, {Bahramian}, {Tingay}, {Prabu}, {Russell}, {Atri}, {Carotenuto}, {Altamirano}, {Motta}, {Hyland}, {Reynolds}, {Weston}, {Fender}, {K{\"o}rding}, {Maitra}, {Markoff}, {Migliari}, {Russell}, {Sarazin}, {Sivakoff}, {Soria}, {Tetarenko}, \& {Tudose}}]{2024ApJ...971L...9W}
{Wood}, C.~M., {Miller-Jones}, J. C.~A., {Bahramian}, A., {et~al.} 2024, \apjl, 971, L9

\bibitem[{{Xu} {et~al.}(2025){Xu}, {You}, {Long}, \& {He}}]{2025arXiv250621131X}
{Xu}, S.-E., {You}, B., {Long}, Y., \& {He}, H. 2025, arXiv e-prints, arXiv:2506.21131

\bibitem[{{Yang} {et~al.}(2024){Yang}, {Zhang}, {Zhang}, {Tao}, {Zhang}, {Ma}, {Bu}, {Huang}, {Liu}, {Yu}, {Xiao}, {Wang}, {Feng}, {Song}, {Ma}, {Ge}, {Zhao}, \& {Qu}}]{2024ApJ...970L..33Y}
{Yang}, Z.-X., {Zhang}, L., {Zhang}, S.-N., {et~al.} 2024, \apjl, 970, L33

\bibitem[{{Young}(2003)}]{2003AdSpR..32.2021Y}
{Young}, A.~J. 2003, Advances in Space Research, 32, 2021

\bibitem[{{Yu} {et~al.}(2024){Yu}, {Bu}, {Zhang}, {Liu}, {Zhang}, {Ducci}, {Tao}, {Santangelo}, {Doroshenko}, {Huang}, {Yang}, \& {Qu}}]{2024MNRAS.529.4624Y}
{Yu}, W., {Bu}, Q.-C., {Zhang}, S.-N., {et~al.} 2024, \mnras, 529, 4624

\bibitem[{{Zhang} {et~al.}(1997){Zhang}, {Cui}, \& {Chen}}]{1997ApJ...482L.155Z}
{Zhang}, S.~N., {Cui}, W., \& {Chen}, W. 1997, \apjl, 482, L155

\bibitem[{{Zhang} {et~al.}(2022){Zhang}, {M{\'e}ndez}, {Garc{\'\i}a}, {Karpouzas}, {Zhang}, {Liu}, {Belloni}, \& {Altamirano}}]{2022MNRAS.514.2891Z}
{Zhang}, Y., {M{\'e}ndez}, M., {Garc{\'\i}a}, F., {et~al.} 2022, \mnras, 514, 2891

\bibitem[{{Zhang} {et~al.}(2024){Zhang}, {M{\'e}ndez}, {Motta}, {Zdziarski}, {Marcel}, {Garc{\'\i}a}, {Altamirano}, {Belloni (deceased)}, {Zhang}, {Timmermans}, \& {Zhang}}]{2024MNRAS.527.5638Z}
{Zhang}, Y., {M{\'e}ndez}, M., {Motta}, S.~E., {et~al.} 2024, \mnras, 527, 5638

\bibitem[{{Zhao} {et~al.}(2024){Zhao}, {Tao}, {Li}, {Zhang}, {Feng}, {Ge}, {Ji}, {Wang}, {Huang}, {Ma}, {Zhang}, {Qu}, {Xu}, {Zhang}, {Yin}, {Shui}, {Ma}, {Zhao}, {Li}, {Yang}, {Liu}, \& {Yu}}]{2024ApJ...961L..42Z}
{Zhao}, Q.-C., {Tao}, L., {Li}, H.-C., {et~al.} 2024, \apjl, 961, L42

\end{thebibliography}
\onecolumn

\begin{appendix} 
\begin{landscape}
\section{Best-fitting parameters of the type-C QPO, $L_\mathrm{l}$ and $L_\mathrm{h}$ observed in the PDS of \Swift. The errors presented in the table represent a $1\sigma$ confidence interval.}
\renewcommand{\arraystretch}{1.7}
\setlength{\tabcolsep}{5pt}

\begin{longtable}{cccccccccccc}
    \hline
    Group ID & Energy (keV) & \multicolumn{3}{c}{Type-C QPO} & \multicolumn{3}{c}{$L_\mathrm{l}$}  & \multicolumn{3}{c}{$L_\mathrm{h}$} 
    \endfirsthead

    \hline
    
    Group ID & Energy (keV) & $f_0$ (Hz) & FWHM (Hz) & norm ($\times 10^{-2}$) &  $f_0$ (Hz) & FWHM (Hz) & norm ($\times 10^{-2}$)  &  $f_0$ (Hz) & FWHM (Hz) & norm ($\times 10^{-2}$) \\ 
    \hline
    \endhead

    \hline
    \endfoot

    \hline
    \endlastfoot
    & & $f_0$ (Hz) & FWHM (Hz) & norm ($\times 10^{-2}$) &  $f_0$ (Hz) & FWHM (Hz) & norm ($\times 10^{-2}$) &  $f_0$ (Hz) & FWHM (Hz) & norm ($\times 10^{-2}$) & \\ \hline

	&	2-4	&	$0.480 \pm 0.002$	&	$0.127 \pm 0.006$	&	$2.39 \pm 0.08$	&	$7.2^{+0.4}_{-0.3}$	&	$4.4^{+1.3}_{-1.1}$	&	$0.10^{+0.04}_{-0.03}$	&	--	&	--	&	--	\\
	&	4-10	&	$0.480 \pm 0.002$	&	$0.122 \pm 0.006$	&	$3.21^{+0.11}_{-0.10}$	&	$7.3^{+0.3}_{-0.4}$	&	$5.2^{+1.4}_{-1.5}$	&	$0.20^{+0.10}_{-0.08}$	&	$0.3^{+2.0}_{-0.3}$	&	$16.1^{+2.1}_{-1.2}$	&	$1.30^{+0.12}_{-0.20}$	\\
	&	10-14	&	$0.513^{+0.003}_{-0.004}$	&	$0.064^{+0.008}_{-0.010}$	&	$3.2 \pm 0.3$	&	$7.7^{+0.4}_{-0.5}$	&	$13.4 \pm 1.2$	&	$0.93 \pm 0.06$	&	$35^{+2}_{-3}$	&	$13^{+7}_{-5}$	&	$0.10 \pm 0.04$	\\
    group1	&	14-23	&	$0.512 \pm 0.003$	&	$0.063^{+0.009}_{-0.006}$	&	$3.4 \pm 0.2$	&	$8.9^{+0.6}_{-0.5}$	&	$13^{+3}_{-2}$	&	$0.66^{+0.11}_{-0.09}$	&	$45^{+11}_{-14}$	&	$60^{+25}_{-16}$	&	$0.22^{+0.07}_{-0.06}$	\\
	&	23-35	&	$0.510^{+0.003}_{-0.004}$	&	$0.055^{+0.010}_{-0.009}$	&	$4.1^{+0.3}_{-0.2}$	&	$5.4^{+1.2}_{-1.9}$	&	$22 \pm 2$	&	$1.16^{+0.20}_{-0.08}$	&	$46^{+7}_{-8}$	&	$77 \pm 12$	&	$0.45^{+0.10}_{-0.09}$	\\
	&	35-50	&	$0.511 \pm 0.004$	&	$0.056 \pm 0.010$	&	$4.6^{+0.3}_{-0.4}$	&	$<0.8$	&	$23^{+2}_{-3}$	&	$2.31^{+0.06}_{-0.13}$	&	$56^{+7}_{-14}$	&	$90^{+18}_{-7}$	&	$0.52^{+0.14}_{-0.08}$	\\
	&	50-100	&	$0.503^{+0.007}_{-0.010}$	&	$0.063^{+0.014}_{-0.018}$	&	$4.6 \pm 1.1$	&	$<4$	&	$21 \pm 2$	&	$3.8^{+0.4}_{-0.5}$	&	$42^{+16}_{-11}$	&	$126^{+16}_{-18}$	&	$1.4^{+0.4}_{-0.3}$	\\

    \hline

	&	2-4	&	$0.702 \pm 0.002$	&	$0.083 \pm 0.004$	&	$1.74 \pm 0.05$	&	$8.9^{+0.4}_{-0.3}$	&	$9.8^{+1.3}_{-1.2}$	&	$0.22^{+0.06}_{-0.03}$	&	$55^{+2}_{-6}$	&	$10^{+8}_{-6}$	&	$0.026^{+0.015}_{-0.011}$	\\
	&	4-10	&	$0.701 \pm 0.002$	&	$0.090^{+0.005}_{-0.006}$	&	$2.79^{+0.11}_{-0.12}$	&	$9.0^{+0.3}_{-0.2}$	&	$11.8^{+1.1}_{-0.7}$	&	$0.62^{+0.08}_{-0.04}$	&	$41^{+3}_{-4}$	&	$41 \pm 9$	&	$0.20 \pm 0.03$	\\
	&	10-14	&	$0.7038^{+0.0013}_{-0.0014}$	&	$0.089^{+0.003}_{-0.004}$	&	$2.84^{+0.07}_{-0.08}$	&	$10.7^{+0.4}_{-0.2}$	&	$6.3^{+2.7}_{-1.4}$	&	$0.22 \pm 0.11$	&	$17^{+4}_{-5}$	&	$23 \pm 5$	&	$0.41^{+0.18}_{-0.13}$	\\
    group2	&	14-23	&	$0.7033^{+0.0013}_{-0.0014}$	&	$0.089^{+0.003}_{-0.004}$	&	$3.10^{+0.07}_{-0.08}$	&	$10.5^{+0.5}_{-0.6}$	&	$10.3^{+2.7}_{-1.0}$	&	$0.39^{+0.09}_{-0.07}$	&	$29^{+3}_{-2}$	&	$33^{+12}_{-10}$	&	$0.30^{+0.07}_{-0.08}$	\\
	&	23-35	&	$0.704 \pm 0.002$	&	$0.090^{+0.002}_{-0.005}$	&	$4.07^{+0.13}_{-0.14}$	&	$9.7^{+1.2}_{-0.8}$	&	$22 \pm 3$	&	$1.0 \pm 0.2$	&	$51^{+8}_{-10}$	&	$89^{+10}_{-12}$	&	$0.58^{+0.11}_{-0.05}$	\\
	&	35-50	&	$0.705 \pm 0.002$	&	$0.086 \pm 0.005$	&	$4.3 \pm 0.2$	&	$6^{+3}_{-4}$	&	$24^{+4}_{-5}$	&	$1.9^{+0.5}_{-0.6}$	&	$50 \pm 20$	&	$109^{+14}_{-13}$	&	$0.9^{+0.4}_{-0.3}$	\\
	&	50-100	&	$0.704 \pm 0.002$	&	$0.088 \pm 0.005$	&	$4.5 \pm 0.2$	&	$9 \pm 2$	&	$16 \pm 5$	&	$2.0^{+1.2}_{-1.0}$	&	$12^{+19}_{-12}$	&	$89^{+19}_{-16}$	&	$2.8^{+0.7}_{-0.8}$	\\

    \hline

	&	2-4	&	$0.862 \pm 0.002$	&	$0.125^{+0.006}_{-0.005}$	&	$1.77 \pm 0.05$	&	$10.2^{+0.3}_{-0.4}$	&	$6.7^{+0.9}_{-1.3}$	&	$0.104^{+0.014}_{-0.024}$	&	$26^{+6}_{-9}$	&	$38^{+12}_{-11}$	&	$0.09^{+0.03}_{-0.02}$	\\
	&	4-10	&	$0.863 \pm 0.002$	&	$0.122 \pm 0.005$	&	$2.67^{+0.06}_{-0.08}$	&	$10.6 \pm 0.2$	&	$9.6^{+0.9}_{-1.1}$	&	$0.35^{+0.05}_{-0.04}$	&	$30^{+3}_{-2}$	&	$42^{+8}_{-4}$	&	$0.22 \pm 0.03$	\\
	&	10-14	&	$0.8595^{+0.0018}_{-0.0010}$	&	$0.125 \pm 0.005$	&	$3.02^{+0.09}_{-0.10}$	&	$11.3^{+0.9}_{-1.2}$	&	$20^{+3}_{-2}$	&	$0.65^{+0.06}_{-0.05}$	&	--	&	--	&	--	\\
    group3	&	14-23	&	$0.861 \pm 0.002$	&	$0.122 \pm 0.005$	&	$3.25^{+0.09}_{-0.10}$	&	$14.5^{+0.6}_{-0.7}$	&	$14^{+3}_{-2}$	&	$0.41^{+0.03}_{-0.05}$	&	$47^{+3}_{-2}$	&	$24^{+10}_{-7}$	&	$0.21^{+0.06}_{-0.05}$	\\
	&	23-35	&	$0.853 \pm 0.002$	&	$0.129 \pm 0.006$	&	$4.02^{+0.13}_{-0.14}$	&	$11 \pm 2$	&	$24^{+4}_{-5}$	&	$0.9 \pm 0.2$	&	$50^{+10}_{-13}$	&	$84^{+7}_{-14}$	&	$0.58^{+0.17}_{-0.08}$	\\
	&	35-50	&	$0.846^{+0.010}_{-0.009}$	&	$0.086^{+0.015}_{-0.018}$	&	$4.0^{+0.8}_{-1.2}$	&	$10.4^{+1.1}_{-2.8}$	&	$21^{+6}_{-5}$	&	$0.9^{+0.4}_{-0.3}$	&	$27^{+6}_{-11}$	&	$78^{+11}_{-8}$	&	$1.1^{+0.4}_{-0.3}$	\\
	&	50-100	&	$0.851^{+0.002}_{-0.003}$	&	$0.135 \pm 0.006$	&	$4.52^{+0.14}_{-0.15}$	&	$8.9^{+0.9}_{-1.7}$	&	$18^{+5}_{-8}$	&	$1.8^{+0.7}_{-1.2}$	&	$17^{+14}_{-17}$	&	$83^{+13}_{-10}$	&	$2.5^{+1.3}_{-0.8}$	\\

    \hline

	&	2-4	&	$1.117 \pm 0.002$	&	$0.158 \pm 0.005$	&	$1.75^{+0.02}_{-0.04}$	&	$12.7 \pm 0.4$	&	$5.3^{+1.6}_{-1.3}$	&	$0.048^{+0.016}_{-0.013}$	&	$19^{+8}_{-14}$	&	$51^{+21}_{-14}$	&	$0.11 \pm 0.03$	\\
	&	4-10	&	$1.116 \pm 0.002$	&	$0.152^{+0.005}_{-0.004}$	&	$2.77^{+0.06}_{-0.05}$	&	$11.1^{+0.7}_{-1.0}$	&	$17.8^{+2.6}_{-1.0}$	&	$0.45 \pm 0.05$	&	$40^{+5}_{-4}$	&	$45^{+14}_{-11}$	&	$0.13 \pm 0.04$	\\
	&	10-14	&	$1.1177 \pm 0.0015$	&	$0.143^{+0.004}_{-0.003}$	&	$2.98 \pm 0.05$	&	$15.9 \pm 0.7$	&	$13^{+2}_{-3}$	&	$0.27 \pm 0.04$	&	$47^{+4}_{-3}$	&	$21^{+6}_{-7}$	&	$0.11^{+0.04}_{-0.02}$	\\
    group4	&	14-23	&	$1.1170^{+0.0014}_{-0.0015}$	&	$0.144^{+0.004}_{-0.002}$	&	$3.26^{+0.07}_{-0.06}$	&	$19.4 \pm 1.0$	&	$22 \pm 3$	&	$0.45^{+0.04}_{-0.05}$	&	$59.0^{+0.7}_{-1.0}$	&	$3^{+4}_{-3}$	&	$0.034^{+0.020}_{-0.014}$	\\
	&	23-35	&	$1.116 \pm 0.002$	&	$0.140 \pm 0.005$	&	$3.86^{+0.09}_{-0.08}$	&	$12 \pm 2$	&	$31 \pm 4$	&	$0.96^{+0.12}_{-0.17}$	&	$70^{+8}_{-9}$	&	$98^{+14}_{-12}$	&	$0.54^{+0.10}_{-0.09}$	\\
	&	35-50	&	$1.115 \pm 0.002$	&	$0.146^{+0.008}_{-0.006}$	&	$4.16^{+0.16}_{-0.11}$	&	$13^{+2}_{-3}$	&	$26^{+5}_{-6}$	&	$1.2^{+0.5}_{-0.4}$	&	$56^{+10}_{-11}$	&	$95 \pm 11$	&	$1.0 \pm 0.2$	\\
	&	50-100	&	$1.117 \pm 0.002$	&	$0.143 \pm 0.006$	&	$4.09^{+0.11}_{-0.12}$	&	$9^{+3}_{-4}$	&	$25 \pm 5$	&	$3.0^{+0.9}_{-1.0}$	&	$46 \pm 14$	&	$101^{+14}_{-13}$	&	$1.8 \pm 0.5$	\\

    \hline

	&	2-4	&	$1.206^{+0.002}_{-0.003}$	&	$0.138 \pm 0.007$	&	$1.38^{+0.04}_{-0.05}$	&	$14.2^{+0.7}_{-1.3}$	&	$15 \pm 3$	&	$0.143^{+0.032}_{-0.011}$	&	--	&	--	&	--	\\
	&	4-10	&	$1.207 \pm 0.002$	&	$0.143^{+0.007}_{-0.006}$	&	$2.30^{+0.07}_{-0.06}$	&	$14.2 \pm 0.6$	&	$6^{+4}_{-2}$	&	$0.07^{+0.04}_{-0.03}$	&	$<12$	&	$50^{+20}_{-30}$	&	$0.54^{+0.10}_{-0.09}$	\\
	&	10-14	&	$1.217 \pm 0.002$	&	$0.135^{+0.007}_{-0.008}$	&	$2.57^{+0.11}_{-0.13}$	&	$15^{+3}_{-4}$	&	$37^{+7}_{-6}$	&	$0.70^{+0.09}_{-0.08}$	&	--	&	--	&	--	\\
    group5	&	14-23	&	$1.220 \pm 0.002$	&	$0.143 \pm 0.006$	&	$2.95^{+0.07}_{-0.08}$	&	$15 \pm 2$	&	$25^{+5}_{-4}$	&	$0.51^{+0.08}_{-0.07}$	&	--	&	--	&	--	\\
	&	23-35	&	$1.218 \pm 0.002$	&	$0.139 \pm 0.007$	&	$3.45 \pm 0.11$	&	$10^{+4}_{-8}$	&	$40^{+11}_{-10}$	&	$1.2 \pm 0.3$	&	$75^{+14}_{-16}$	&	$60^{+20}_{-40}$	&	$0.3 \pm 0.2$	\\
	&	35-50	&	$1.218 \pm 0.002$	&	$0.136^{+0.007}_{-0.006}$	&	$3.56 \pm 0.12$	&	$12^{+4}_{-7}$	&	$29 \pm 7$	&	$1.4^{+0.7}_{-0.8}$	&	$60 \pm 20$	&	$110 \pm 20$	&	$0.8^{+0.4}_{-0.3}$	\\
	&	50-100	&	$1.219^{+0.002}_{-0.003}$	&	$0.154^{+0.008}_{-0.007}$	&	$3.96 \pm 0.12$	&	$10^{+3}_{-7}$	&	$25^{+11}_{-15}$	&	$<4$	&	$40^{+20}_{-30}$	&	$90 \pm 20$	&	$1.7^{+1.6}_{-0.9}$	\\

    \hline

	&	2-4	&	$1.105 \pm 0.002$	&	$0.152^{+0.006}_{-0.003}$	&	$1.45 \pm 0.04$	&	$14.0 \pm 0.6$	&	$7 \pm 2$	&	$0.058 \pm 0.015$	&	$39 \pm 6$	&	$32^{+17}_{-6}$	&	$0.06 \pm 0.02$	\\
	&	4-10	&	$1.105 \pm 0.002$	&	$0.137 \pm 0.005$	&	$2.22 \pm 0.05$	&	$13.5^{+0.4}_{-0.5}$	&	$9 \pm 2$	&	$0.18 \pm 0.04$	&	$32^{+4}_{-6}$	&	$45^{+11}_{-9}$	&	$0.23^{+0.05}_{-0.04}$	\\
	&	10-14	&	$1.1081^{+0.0010}_{-0.0019}$	&	$0.139 \pm 0.006$	&	$2.38 \pm 0.07$	&	$16.5^{+1.4}_{-1.7}$	&	$24^{+4}_{-3}$	&	$0.61^{+0.09}_{-0.07}$	&	--	&	--	&	--	\\
    group6	&	14-23	&	$1.106 \pm 0.002$	&	$0.135^{+0.003}_{-0.005}$	&	$2.56 \pm 0.06$	&	$18^{+2}_{-3}$	&	$23^{+7}_{-5}$	&	$0.39^{+0.09}_{-0.07}$	&	--	&	--	&	--	\\
	&	23-35	&	$1.100 \pm 0.002$	&	$0.143 \pm 0.006$	&	$3.33^{+0.08}_{-0.09}$	&	$15.8^{+1.0}_{-1.5}$	&	$30^{+5}_{-6}$	&	$0.8 \pm 0.2$	&	$68^{+7}_{-17}$	&	$70 \pm 20$	&	$0.43^{+0.17}_{-0.12}$	\\
	&	35-50	&	$1.100 \pm 0.002$	&	$0.129 \pm 0.006$	&	$3.24^{+0.08}_{-0.05}$	&	$12^{+2}_{-4}$	&	$28 \pm 6$	&	$1.4 \pm 0.4$	&	$68^{+9}_{-14}$	&	$80 \pm 20$	&	$0.6 \pm 0.2$	\\
	&	50-100	&	$1.102 \pm 0.002$	&	$0.129 \pm 0.007$	&	$3.27 \pm 0.12$	&	$8^{+4}_{-6}$	&	$30^{+4}_{-5}$	&	$3.3^{+0.8}_{-0.9}$	&	$60^{+11}_{-13}$	&	$80 \pm 20$	&	$1.0 \pm 0.4$	\\

    \hline

	&	2-4	&	$1.337 \pm 0.003$	&	$0.216 \pm 0.008$	&	$1.34 \pm 0.03$	&	$14^{+3}_{-8}$	&	$26^{+11}_{-9}$	&	$0.12^{+0.05}_{-0.04}$	&	--	&	--	&	--	\\
	&	4-10	&	$1.335 \pm 0.003$	&	$0.209 \pm 0.008$	&	$2.12 \pm 0.06$	&	$14.0^{+1.5}_{-2.2}$	&	$20 \pm 5$	&	$0.30^{+0.10}_{-0.09}$	&	$52^{+13}_{-14}$	&	$90^{+20}_{-30}$	&	$0.23^{+0.11}_{-0.07}$	\\
	&	10-14	&	$1.340 \pm 0.002$	&	$0.200 \pm 0.005$	&	$2.43 \pm 0.04$	&	$17.7^{+1.3}_{-1.7}$	&	$16^{+8}_{-5}$	&	$0.26^{+0.11}_{-0.05}$	&	$42^{+6}_{-3}$	&	$33^{+12}_{-10}$	&	$0.22^{+0.09}_{-0.08}$	\\
    group7	&	14-23	&	$1.317^{+0.007}_{-0.006}$	&	$0.12 \pm 0.02$	&	$2.3 \pm 0.3$	&	$21^{+3}_{-4}$	&	$32^{+11}_{-8}$	&	$0.39^{+0.08}_{-0.07}$	&	--	&	--	&	--	\\
	&	23-35	&	$1.331 \pm 0.002$	&	$0.206 \pm 0.006$	&	$3.09 \pm 0.06$	&	$17.2^{+1.0}_{-1.4}$	&	$22^{+6}_{-5}$	&	$0.48^{+0.13}_{-0.11}$	&	$59^{+8}_{-9}$	&	$89^{+13}_{-12}$	&	$0.71 \pm 0.11$	\\
	&	35-50	&	$1.330 \pm 0.002$	&	$0.205^{+0.007}_{-0.006}$	&	$3.32 \pm 0.07$	&	$<8$	&	$49^{+4}_{-3}$	&	$1.9 \pm 0.2$	&	--	&	--	&	--	\\
	&	50-100	&	$1.331 \pm 0.003$	&	$0.208 \pm 0.008$	&	$3.38 \pm 0.09$	&	$13.4^{+0.8}_{-1.0}$	&	$15^{+6}_{-5}$	&	$1.1^{+0.6}_{-0.5}$	&	$<36$	&	$100 \pm 20$	&	$2.5 \pm 0.6$	\\
    \end{longtable}
\end{landscape}

\begin{landscape}
\section{The characteristic frequency and fractional rms of the type-C QPO, $L_\mathrm{l}$ and $L_\mathrm{h}$, together with the phase-lag of the type-C QPO and $L_\mathrm{l}$. The errors presented in the table represent a $1\sigma$ confidence interval.}
\renewcommand{\arraystretch}{1.6}
\setlength{\tabcolsep}{7pt}
\begin{longtable}{cccccccccc}
    \label{tab:longtable2} \\
    \hline
    Group ID & Energy (keV) & \multicolumn{3}{c}{Type-C QPO} & \multicolumn{3}{c}{$L_\mathrm{l}$}  & \multicolumn{2}{c}{$L_\mathrm{h}$}\\ 
    & & $\nu_{max}$ (Hz) & RMS ($\%$) & Phase lag (rad) & $\nu_{max}$ (Hz) & RMS ($\%$) & Phase lag (rad) & $\nu_{max}$ (Hz) & RMS ($\%$) \\ \hline
    \endfirsthead

    \hline
    Group ID & Energy (keV) & $\nu_{max}$ (Hz) & RMS ($\%$) & Phase lag (rad) & $\nu_{max}$ (Hz) & RMS ($\%$) & Phase lag (rad) & $\nu_{max}$ (Hz) & RMS ($\%$) \\ \hline
    \endhead

    \hline
    \endfoot

    \hline
    \endlastfoot

	&	2-4	&	$0.497 \pm 0.004$	&	$15.5^{+0.2}_{-0.3}$	&	--	&	$8.4^{+1.0}_{-0.8}$	&	$3.2^{+0.7}_{-0.5}$	&	--	&	--	&	--	\\
	&	4-10	&	$0.495 \pm 0.004$	&	$17.9 \pm 0.3$	&	$0.06 \pm 0.03$	&	$9.0^{+1.0}_{-1.2}$	&	$4.5^{+1.1}_{-0.9}$	&	$-0.11 \pm 0.13$	&	$16.1^{+2.1}_{-1.2}$	&	$0.114^{+0.005}_{-0.009}$	\\
	&	10-14	&	$0.517^{+0.004}_{-0.005}$	&	$17.8 \pm 0.7$	&	$0.10 \pm 0.04$	&	$15.5 \pm 1.2$	&	$9.6 \pm 0.3$	&	$-0.05 \pm 0.04$	&	$37^{+5}_{-4}$	&	$0.031 \pm 0.006$	\\
    group1	&	14-23	&	$0.516 \pm 0.004$	&	$18.3^{+0.7}_{-0.5}$	&	$0.09 \pm 0.05$	&	$16^{+3}_{-2}$	&	$8.1^{+0.7}_{-0.6}$	&	$0.07 \pm 0.07$	&	$80^{+30}_{-20}$	&	$0.047^{+0.008}_{-0.006}$	\\
	&	23-35	&	$0.513^{+0.004}_{-0.005}$	&	$20.2^{+0.8}_{-0.6}$	&	$0.07 \pm 0.07$	&	$23 \pm 2$	&	$10.8^{+0.9}_{-0.4}$	&	$-0.30 \pm 0.10$	&	$89 \pm 14$	&	$0.067 \pm 0.007$	\\
	&	35-50	&	$0.514 \pm 0.005$	&	$21.5^{+0.7}_{-1.0}$	&	$0.07 \pm 0.07$	&	$23^{+2}_{-3}$	&	$15.2^{+0.2}_{-0.4}$	&	$-0.04^{+0.35}_{-0.28}$	&	$106^{+19}_{-14}$	&	$0.072^{+0.010}_{-0.005}$	\\
	&	50-100	&	$0.507^{+0.009}_{-0.012}$	&	$22 \pm 3$	&	$0.03 \pm 0.09$	&	$21 \pm 2$	&	$19.5^{+1.0}_{-1.2}$	&	$-1.0 \pm 0.3$	&	$130 \pm 20$	&	$0.117^{+0.017}_{-0.015}$	\\

    \hline

	&	2-4	&	$0.707 \pm 0.002$	&	$13.2 \pm 0.2$	&	--	&	$13.2^{+1.2}_{-1.1}$	&	$4.7^{+0.6}_{-0.3}$	&	--	&	$56^{+3}_{-7}$	&	$0.016^{+0.005}_{-0.003}$	\\
	&	4-10	&	$0.707 \pm 0.003$	&	$16.7^{+0.3}_{-0.4}$	&	$0.05 \pm 0.04$	&	$14.8^{+1.0}_{-0.7}$	&	$7.9^{+0.5}_{-0.3}$	&	$-0.04 \pm 0.04$	&	$58 \pm 9$	&	$0.044 \pm 0.004$	\\
	&	10-14	&	$0.709 \pm 0.002$	&	$16.9 \pm 0.2$	&	$0.17 \pm 0.04$	&	$12.5^{+1.7}_{-0.9}$	&	$4.7 \pm 1.1$	&	$-0.07^{+0.25}_{-0.27}$	&	$29 \pm 7$	&	$0.064^{+0.014}_{-0.010}$	\\
    group2	&	14-23	&	$0.709 \pm 0.002$	&	$17.6 \pm 0.2$	&	$0.16 \pm 0.05$	&	$14.7^{+2.2}_{-1.1}$	&	$6.3^{+0.7}_{-0.5}$	&	$-0.3 \pm 0.2$	&	$44^{+11}_{-9}$	&	$0.055^{+0.006}_{-0.007}$	\\
	&	23-35	&	$0.710 \pm 0.002$	&	$20.2^{+0.3}_{-0.4}$	&	$0.16 \pm 0.06$	&	$24 \pm 3$	&	$9.9^{+0.9}_{-0.8}$	&	$-0.3 \pm 0.2$	&	$103^{+12}_{-16}$	&	$0.076^{+0.007}_{-0.004}$	\\
	&	35-50	&	$0.710 \pm 0.002$	&	$20.6 \pm 0.4$	&	$0.16 \pm 0.06$	&	$25^{+4}_{-6}$	&	$14 \pm 2$	&	$0.09 \pm 0.22$	&	$120 \pm 20$	&	$0.10 \pm 0.02$	\\
	&	50-100	&	$0.710 \pm 0.002$	&	$21.3 \pm 0.4$	&	$0.14 \pm 0.06$	&	$18^{+6}_{-5}$	&	$14 \pm 4$	&	$-0.5^{+1.6}_{-0.6}$	&	$90^{+21}_{-17}$	&	$0.17^{+0.02}_{-0.03}$	\\

    \hline

	&	2-4	&	$0.871 \pm 0.003$	&	$13.3 \pm 0.2$	&	--	&	$12.2^{+0.8}_{-1.0}$	&	$3.2^{+0.2}_{-0.4}$	&	--	&	$46^{+13}_{-14}$	&	$0.031^{+0.005}_{-0.004}$	\\
	&	4-10	&	$0.871 \pm 0.003$	&	$16.3 \pm 0.2$	&	$0.07 \pm 0.04$	&	$14.3^{+0.7}_{-0.8}$	&	$5.9 \pm 0.4$	&	$-0.07 \pm 0.06$	&	$52^{+9}_{-5}$	&	$0.046^{+0.003}_{-0.004}$	\\
	&	10-14	&	$0.869^{+0.003}_{-0.002}$	&	$17.4 \pm 0.3$	&	$0.11 \pm 0.03$	&	$23 \pm 3$	&	$8.1^{+0.4}_{-0.3}$	&	$-0.14 \pm 0.08$	&	--	&	--	\\
    group3	&	14-23	&	$0.869 \pm 0.003$	&	$18.0 \pm 0.3$	&	$0.14 \pm 0.04$	&	$20 \pm 2$	&	$6.4^{+0.2}_{-0.4}$	&	$-0.29^{+0.13}_{-0.12}$	&	$53^{+7}_{-5}$	&	$0.046^{+0.006}_{-0.005}$	\\
	&	23-35	&	$0.863 \pm 0.003$	&	$20.0 \pm 0.3$	&	$0.10 \pm 0.06$	&	$27^{+4}_{-6}$	&	$9.3^{+1.3}_{-1.2}$	&	$-0.8^{+0.5}_{-0.8}$	&	$98^{+11}_{-19}$	&	$0.076^{+0.012}_{-0.005}$	\\
	&	35-50	&	$0.851^{+0.011}_{-0.010}$	&	$20^{+2}_{-3}$	&	$0.007^{+0.086}_{-0.087}$	&	$24 \pm 6$	&	$10 \pm 2$	&	$-0.04^{+0.25}_{-0.24}$	&	$82^{+12}_{-11}$	&	$0.106^{+0.017}_{-0.014}$	\\
	&	50-100	&	$0.862^{+0.003}_{-0.004}$	&	$21.3 \pm 0.3$	&	$0.07 \pm 0.06$	&	$20^{+5}_{-8}$	&	$14^{+3}_{-4}$	&	$0.04^{+0.44}_{-0.20}$	&	$85^{+16}_{-14}$	&	$0.16^{+0.04}_{-0.02}$	\\

    \hline

	&	2-4	&	$1.128 \pm 0.002$	&	$13.23^{+0.07}_{-0.14}$	&	--	&	$13.8^{+0.9}_{-0.8}$	&	$2.2^{+0.4}_{-0.3}$	&	--	&	$50 \pm 20$	&	$0.034^{+0.005}_{-0.004}$	\\
	&	4-10	&	$1.126 \pm 0.002$	&	$16.6 \pm 0.2$	&	$0.06 \pm 0.02$	&	$21.0^{+2.5}_{-1.4}$	&	$6.7^{+0.3}_{-0.4}$	&	$-0.11^{+0.04}_{-0.05}$	&	$60^{+14}_{-10}$	&	$0.036 \pm 0.005$	\\
	&	10-14	&	$1.127 \pm 0.002$	&	$17.26^{+0.15}_{-0.13}$	&	$0.03 \pm 0.02$	&	$21 \pm 2$	&	$5.2 \pm 0.4$	&	$-0.44 \pm 0.08$	&	$51 \pm 6$	&	$0.033^{+0.005}_{-0.002}$	\\
    group4	&	14-23	&	$1.126 \pm 0.002$	&	$18.1 \pm 0.2$	&	$0.04 \pm 0.02$	&	$29 \pm 3$	&	$6.7^{+0.3}_{-0.4}$	&	$-0.51^{+0.10}_{-0.09}$	&	$59.0^{+0.9}_{-1.2}$	&	$0.018^{+0.006}_{-0.004}$	\\
	&	23-35	&	$1.124 \pm 0.002$	&	$19.6 \pm 0.2$	&	$0.06 \pm 0.03$	&	$33^{+5}_{-4}$	&	$9.8^{+0.6}_{-0.9}$	&	$-0.37^{+0.14}_{-0.13}$	&	$120 \pm 15$	&	$0.074^{+0.007}_{-0.006}$	\\
	&	35-50	&	$1.125 \pm 0.003$	&	$20.4^{+0.4}_{-0.3}$	&	$0.06 \pm 0.03$	&	$29 \pm 6$	&	$11 \pm 2$	&	$-0.26^{+0.12}_{-0.11}$	&	$110 \pm 15$	&	$0.098^{+0.009}_{-0.011}$	\\
	&	50-100	&	$1.126 \pm 0.003$	&	$20.2 \pm 0.3$	&	$0.04 \pm 0.03$	&	$27 \pm 6$	&	$17 \pm 3$	&	$0.3^{+0.3}_{-0.4}$	&	$111^{+19}_{-17}$	&	$0.14 \pm 0.02$	\\

    \hline

	&	2-4	&	$1.214 \pm 0.003$	&	$11.7 \pm 0.2$	&	--	&	$21 \pm 3$	&	$3.79^{+0.42}_{-0.15}$	&	--	&	--	&	--	\\
	&	4-10	&	$1.215 \pm 0.003$	&	$15.2 \pm 0.2$	&	$0.08 \pm 0.04$	&	$16 \pm 2$	&	$2.6^{+0.9}_{-0.5}$	&	$-0.01 \pm 0.38$	&	$50^{+20}_{-30}$	&	$0.074^{+0.007}_{-0.006}$	\\
	&	10-14	&	$1.225 \pm 0.003$	&	$16.0^{+0.3}_{-0.4}$	&	$-0.10 \pm 0.05$	&	$40 \pm 7$	&	$8.4 \pm 0.5$	&	$-0.30 \pm 0.15$	&	--	&	--	\\
    group5	&	14-23	&	$1.228 \pm 0.003$	&	$17.2 \pm 0.2$	&	$-0.07 \pm 0.03$	&	$29^{+5}_{-4}$	&	$7.2 \pm 0.5$	&	$-0.6 \pm 0.2$	&	--	&	--	\\
	&	23-35	&	$1.226 \pm 0.003$	&	$18.6 \pm 0.3$	&	$-0.05 \pm 0.04$	&	$41^{+11}_{-12}$	&	$11 \pm 2$	&	$-0.9 \pm 0.2$	&	$100^{+30}_{-40}$	&	$0.06 \pm 0.02$	\\
	&	35-50	&	$1.226 \pm 0.003$	&	$18.9 \pm 0.3$	&	$-0.06^{+0.04}_{-0.03}$	&	$31^{+8}_{-9}$	&	$12 \pm 3$	&	$0.2^{+0.6}_{-0.5}$	&	$120 \pm 30$	&	$0.09 \pm 0.02$	\\
	&	50-100	&	$1.228 \pm 0.003$	&	$19.9 \pm 0.3$	&	$-0.04 \pm 0.04$	&	$27^{+11}_{-17}$	&	$15 \pm 5$	&	$0.13^{+0.49}_{-0.48}$	&	$100 \pm 30$	&	$0.13^{+0.06}_{-0.04}$	\\

    \hline

	&	2-4	&	$1.115 \pm 0.003$	&	$12.1 \pm 0.2$	&	--	&	$15.5^{+1.4}_{-1.2}$	&	$2.4 \pm 0.3$	&	--	&	$50^{+16}_{-9}$	&	$0.025 \pm 0.004$	\\
	&	4-10	&	$1.113 \pm 0.003$	&	$14.9 \pm 0.2$	&	$0.08 \pm 0.03$	&	$16.1^{+1.4}_{-1.3}$	&	$4.2 \pm 0.5$	&	$-0.05 \pm 0.19$	&	$55 \pm 11$	&	$0.048^{+0.005}_{-0.004}$	\\
	&	10-14	&	$1.117^{+0.002}_{-0.003}$	&	$15.4 \pm 0.2$	&	$0.07 \pm 0.03$	&	$29 \pm 4$	&	$7.8 \pm 0.5$	&	$-0.15^{+0.12}_{-0.13}$	&	--	&	--	\\
    group6	&	14-23	&	$1.114 \pm 0.002$	&	$16.0 \pm 0.2$	&	$0.08^{+0.03}_{-0.02}$	&	$29 \pm 6$	&	$6.2^{+0.7}_{-0.5}$	&	$-0.22 \pm 0.13$	&	--	&	--	\\
	&	23-35	&	$1.109 \pm 0.003$	&	$18.2 \pm 0.2$	&	$0.07 \pm 0.03$	&	$34^{+5}_{-6}$	&	$9.1^{+0.9}_{-1.1}$	&	$-0.40^{+0.14}_{-0.13}$	&	$100 \pm 20$	&	$0.066^{+0.013}_{-0.009}$	\\
	&	35-50	&	$1.108^{+0.002}_{-0.003}$	&	$18.00^{+0.21}_{-0.14}$	&	$0.05 \pm 0.03$	&	$31^{+6}_{-7}$	&	$12 \pm 2$	&	$-0.4^{+0.2}_{-0.3}$	&	$100 \pm 20$	&	$0.080^{+0.015}_{-0.012}$	\\
	&	50-100	&	$1.110 \pm 0.003$	&	$18.1 \pm 0.3$	&	$0.05^{+0.04}_{-0.03}$	&	$31^{+5}_{-6}$	&	$18^{+2}_{-3}$	&	$-1.4^{+0.4}_{-0.2}$	&	$100 \pm 30$	&	$0.10 \pm 0.02$	\\

    \hline

	&	2-4	&	$1.354 \pm 0.004$	&	$11.56 \pm 0.15$	&	--	&	$30 \pm 11$	&	$3.5^{+0.7}_{-0.5}$	&	--	&	--	&	--	\\
	&	4-10	&	$1.351 \pm 0.004$	&	$14.6 \pm 0.2$	&	$0.06 \pm 0.02$	&	$24 \pm 5$	&	$5.4 \pm 0.9$	&	$-0.16^{+0.13}_{-0.14}$	&	$100^{+20}_{-30}$	&	$0.048^{+0.011}_{-0.008}$	\\
	&	10-14	&	$1.354 \pm 0.003$	&	$15.58 \pm 0.13$	&	$0.06 \pm 0.02$	&	$24^{+6}_{-5}$	&	$5.1^{+1.0}_{-0.4}$	&	$-0.41 \pm 0.12$	&	$53^{+13}_{-9}$	&	$0.047^{+0.009}_{-0.008}$	\\
    group7	&	14-23	&	$1.322 \pm 0.008$	&	$15.2^{+1.1}_{-1.0}$	&	$0.06 \pm 0.03$	&	$38^{+11}_{-9}$	&	$6.2^{+0.7}_{-0.6}$	&	$-0.3^{+0.6}_{-0.3}$	&	--	&	--	\\
	&	23-35	&	$1.347 \pm 0.003$	&	$17.6 \pm 0.2$	&	$0.08 \pm 0.02$	&	$28 \pm 5$	&	$6.9^{+1.0}_{-0.8}$	&	$-0.71^{+0.13}_{-0.12}$	&	$107 \pm 15$	&	$0.084 \pm 0.007$	\\
	&	35-50	&	$1.346 \pm 0.003$	&	$18.2 \pm 0.2$	&	$0.08 \pm 0.02$	&	$49 \pm 4$	&	$13.7^{+0.6}_{-0.7}$	&	$-1.0 \pm 0.3$	&	--	&	--	\\
	&	50-100	&	$1.347 \pm 0.004$	&	$18.4^{+0.2}_{-0.3}$	&	$0.09 \pm 0.03$	&	$20^{+5}_{-4}$	&	$11^{+3}_{-2}$	&	$0.8^{+0.2}_{-0.3}$	&	$100 \pm 20$	&	$0.16 \pm 0.02$	\\
\end{longtable}
    
\end{landscape}

\begin{table*}
\label{A.C}
\section{The best-fitting spectral parameters fitting to the NuSTAR spectrum (ObsID: 80902333004) of \Swift\ with the model \texttt{Constant*Tbabs(diskbb+relxill+cutoffpl)}. 
Uncertainties are given at the 90 percent confidence level. The parameters $N_{\rm H}$, $a_*$ and emissivity indices ($\alpha_1$, $\alpha_2$) have been fixed during the fitting. }
        \centering
            \label{Table1}
            \begin{tabular}{ccc}
                \hline
                \noalign{\smallskip}
                Component & Parameter & Value \\
                \noalign{\smallskip}
                \hline
                \noalign{\smallskip}
                TBabs  &  $N_{\rm H}\ (10^{20}\ \rm cm^{-2})$  & [0.3] \\
                \noalign{\smallskip}
                \hline
                \noalign{\smallskip}
                diskbb & $\rm kT_{in}\  (keV)$ & $0.29 \pm 0.03$\\
                \noalign{\smallskip}
                & $\rm N_{disk}\ (10^6)$ & $4.2_{-2.6}^{+10.3}$ \\
                \noalign{\smallskip}
                \hline
                \noalign{\smallskip}
                relxill & $\alpha_1,\ \alpha_2$ & [3] \\
                \noalign{\smallskip}
                & $a_*$ & [0.2] \\
                \noalign{\smallskip}
                & $\rm R_{in}\ (R_{ISCO})$ & $1.7_{-0.5}^{+1.1}$ \\
                \noalign{\smallskip}
                & $i \ (\degree)$ & $30_{-6}^{+3}$ \\
                \noalign{\smallskip}
                & $\Gamma_1$ & $1.95_{-0.07}^{+0.10}$ \\
                \noalign{\smallskip}
                & $E_{\rm cut} \ (\rm keV)$ & $72_{-17}^{+54}$ \\
                \noalign{\smallskip}
                & $\log \xi \ (\log \ [\rm erg\ cm \ s^{-1}])$ & $3.41_{-0.16}^{+0.26}$ \\
                \noalign{\smallskip}
                & $A_{\rm Fe}$ & $0.9_{-0.3}^{+1.4}$ \\
                \noalign{\smallskip}
                & $refl_{\rm frac}$ & $0.086_{-0.018}^{+0.033}$ \\
                \noalign{\smallskip}
                & $N_{\rm relxill}$ & $0.31_{-0.04}^{+0.03}$ \\
                \noalign{\smallskip}
                \hline
                \noalign{\smallskip}
                cutoffpl & $\Gamma_2$ & $1.16_{-0.07}^{+0.08}$ \\
                \noalign{\smallskip}
                & $E_{\rm cut}\ (\rm keV)$ & $11.7_{-0.6}^{+1.2}$ \\
                \noalign{\smallskip}
                & $N_{\rm cutoffpl}$ & $9.2_{-1.3}^{+2.6}$ \\
                \noalign{\smallskip}
                \hline
                \noalign{\smallskip}
                & $\chi^2 /d.o.f$ & 2285.23/2183 \\
                \noalign{\smallskip}
                \hline
            \end{tabular}
    \end{table*}

\end{appendix}
\end{document}